\pdfoutput=1
\documentclass[bib]{statapress}

\usepackage[crop,newcenter,frame]{pagedims}

\usepackage{bm} 
\usepackage{bbm}
\usepackage{amsmath}
\usepackage{amssymb}
\usepackage{amsfonts}
\usepackage{subcaption}
\usepackage{float}
\usepackage{enumitem}
\usepackage{multirow}
\usepackage{tabularx}
\usepackage{xspace}
\usepackage{xcolor}
\usepackage{booktabs}
\usepackage{setspace}
\usepackage[normalem]{ulem} 
\usepackage{pifont}
\usepackage{threeparttable} 
\usepackage{rotating}

\usepackage{xr}
\newtheorem{assumption}{Assumption}

\usepackage{sj}
\usepackage{epsfig}
\usepackage{stata}
\usepackage{shadow}
\sjsetissue{$vv$}{$ii$}{$mm$}{$yyyy$}

\input{partialinput}

\begin{document}


\graphicspath{{fig/}}

\makeatletter
\def\input@path{{tab/}{mc/}}
\makeatother

\hypersetup{hidelinks}


\newcommand{\sjversion}[2]{#1}

\newcommand{\norm}[1]{\left\lVert#1\right\rVert}
\newcommand{\normline}[1]{\lVert#1\rVert}

\newcommand{\myoption}[2]{\texttt{#1(}#2\texttt{)}}
\newcommand{\ttt}[1]{\texttt{#1}}
\newcommand{\lassopack}{\stcmd{lassopack}\xspace}
\newcommand{\pdslasso}{\stcmd{pdslasso}\xspace}
\newcommand{\lassotwo}{\stcmd{lasso2}\xspace}
\newcommand{\cvlasso}{\stcmd{cvlasso}\xspace}
\newcommand{\rlasso}{\stcmd{rlasso}\xspace}
\newcommand{\lasso}{lasso\xspace}
\newcommand{\ddml}{\stcmd{ddml}\xspace}
\newcommand{\qddml}{\stcmd{qddml}\xspace}
\newcommand{\pystacked}{\stcmd{pystacked}\xspace}
\newcommand{\pylasso}{\stcmd{pylasso}\xspace}
\newcommand{\crossfit}{\stcmd{crossfit}\xspace}

\newcommand{\stopt}[2]{\texttt{#1(}\emph{#2}\texttt{)}}

\newcommand{\real}{{\it real}}
\newcommand{\stint}{{\it integer}}
\newcommand{\stvarname}{{\it varname}}
\newcommand{\var}{{\it variable}}
\newcommand{\method}{{\it method}}
\newcommand{\stnumlist}{{\it numlist}}
\newcommand{\stmatrix}{{\it matrix}}

\newcommand{\achim}[1]{\textcolor{red}{$<<$AA: #1$>>$}}
\newcommand{\ms}[1]{\textcolor{blue}{$<<$MS: #1$>>$}}
\newcommand{\chris}[1]{\textcolor{violet}{$<<$CH: #1$>>$}}

\newcommand{\aarevised}[2]{\textcolor{red}{\sout{#1} #2}}
\newcommand{\msrevised}[2]{\textcolor{blue}{\sout{#1} #2}}

\newcommand{\revised}[1]{#1} 
\newcommand{\revisedcolor}{}
\newcommand{\revisedcoloroff}{}


\inserttype[st0001]{article}
\author{Ahrens, Hansen, Schaffer \& Wiemann}{%
  Achim Ahrens\\ETH Z\"urich \\achim.ahrens@gess.ethz.ch
  \and
  Christian B. Hansen\\  University of Chicago\\christian.hansen@chicagobooth.edu
  \vspace{.3cm}
  \and
  Mark E. Schaffer\\ Heriot-Watt University\\Edinburgh, United Kingdom \\ m.e.schaffer@hw.ac.uk
  \and
  Thomas Wiemann\\  University of Chicago\\ wiemann@uchicago.edu
}
\title[ddml]{ddml: Double/debiased machine learning \\in Stata}

\maketitle
\sjversion{}{\thispagestyle{empty}}

\begin{abstract}
We introduce the package \ddml for Double/De\-bias\-ed Machine Learning (DDML) in Stata. Estimators of causal parameters for five different econometric models are supported, allowing for flexible estimation of causal effects of endogenous variables in settings with unknown functional forms and/or many exogenous variables. \ddml is compatible with many existing supervised machine learning programs in Stata. We recommend using DDML in combination with stacking estimation which combines multiple machine learners into a final predictor. We provide Monte Carlo evidence to support our recommendation. 

\end{abstract}

\section{Introduction}

Identification of causal effects frequently relies on an unconfoundedness assumption, requiring that treatment or instrument assignment is sufficiently random given observed control covariates. Estimation of causal effects in these settings then involves conditioning on the controls. Unfortunately, estimators of causal effects that are insufficiently flexible to capture the effect of confounds generally do not produce consistent estimates of causal effects even when unconfoundedness holds. For example, \citet{blandhol2022tsls} highlight that TSLS estimands obtained after controlling linearly for confounds do not generally correspond to weakly causal effects even when instruments are valid conditional on controls. Even in the ideal scenario where theory provides a small number of relevant controls, theory rarely specifies the exact nature of confounding. Thus, applied empirical researchers wishing to exploit unconfoundedness assumptions to learn causal effects face a nonparametric estimation problem. 

Traditional nonparametric estimators suffer greatly under the curse of dimensionality and are quickly impractical in the frequently encountered setting with multiple observed covariates.\footnote{For example, the number of coefficients in polynomial series regression with interaction terms increases exponentially in the number of covariates.} These difficulties leave traditional nonparametric estimators essentially inapplicable in the presence of increasingly large and complex data sets, e.g.\ textual confounders as in \citet{roberts2020a} or digital trace data \citep{hangartner2021a}. Tools from supervised machine learning have been put forward as alternative estimators. These approaches are often more robust to the curse of dimensionality via the exploitation of regularization assumptions. A prominent example of a machine learning-based causal effects estimator is Post-Double Selection Lasso (PDS-Lasso) of \citet{Belloni2014a}, which fits auxiliary lasso regressions of the outcome and treatment(s), respectively, against a menu of transformed controls. Under an approximate sparsity assumption, which posits that the DGP can be approximated well by a relatively small number of terms included in the menu, this approach allows for precise treatment effect estimation. The lasso can also be used for approximating optimal instruments \citep{Belloni2012}. Lasso-based approaches for estimation of causal effects have become a popular strategy in applied econometrics \citep[e.g.][]{gilchrist2016something,dhar2022reshaping}, partially facilitated by the availability of software programs in Stata (\pdslasso, \citealp{Ahrens2018}; \citealp{statacorp2019}) and R (\texttt{hdm}, \citealp{Chernozhukov2016b}).

Although approximate sparsity is a weaker regularization assumption than assuming a linear functional form that depends on a known low-dimensional set of variables, it may not be suitable in a wide range of applications. For example, \citet{giannone2021economic} argue that approximate sparsity may provide a poor description in several economic examples. There is thus a potential benefit to expanding the set of regularization assumptions and correspondingly considering a larger set of machine learners including, for example, random forests, gradient boosting, and neural networks. While the theoretical properties of these estimators are an active research topic (see, e.g., \citealp{Athey2019a, farrell2021deep}), machine learning methods are widely adopted in industry and practice for their empirical performance. To facilitate their application for causal inference in common econometric models, \citet{Chernozhukov2018} propose Double/Debiased Machine Learning (DDML), which exploits Neyman orthogonality of estimating equations and cross-fitting to formally establish asymptotic normality of estimators of causal parameters under relatively mild convergence rate conditions on nonparametric estimators.

DDML increases the set of machine learners that researchers can leverage for estimation of causal effects. Deciding which learner is most suitable for a particular application is difficult, however, since researchers are rarely certain about the structure of the underlying data generating process. A practical solution is to construct combinations of a diverse set of machine learners using stacking \citep{Wolpert1992,Breiman1996a}. Stacking is a meta learner given by a weighted sum of individual machine learners (the ``base learners''). When the weights corresponding to the base learners are chosen to maximize out-of-sample predictive accuracy, this approach hedges against the risk of relying on any particular poorly suited or ill-tuned machine learner.

In this article, we introduce the Stata package \ddml, which implements DDML for Stata.\footnote{This article refers to version \revised{v1.4.2} of \ddml.} \ddml adds to a small number of programs for causal machine learning in Stata (\citealp{Ahrens2018}, \citealp{statacorp2019}, \citealp{statacorp2021}). We briefly summarize the four main features of the program: 
\begin{enumerate}

\item \ddml supports flexible estimators of causal parameters in five econometric models: (1) the Partially Linear Model, (2) the Interactive Model (for binary treatment), (3) the Partially Linear IV Model, (4) the Flexible Partially Linear IV Model, and (5) the Interactive IV Model (for binary treatment and instrument).

\item \ddml supports data-driven combinations of multiple machine learners via stacking by leveraging \pystacked \citep{Ahrens2022}, our complementary Stata frontend relying on the Python library \emph{scikit-learn} \citep{scikit-learn,sklearn_api}. \revised{\ddml also supports two novel approaches to pairing DDML with stacking introduced in \citet{Ahrens2023_applied}: short-stacking takes a short-cut by leveraging the cross-fitted predicted values for estimating the stacking weights and pooled stacking enforces common weights across cross-fitting folds.}

\item Aside from \pystacked, \ddml can be used in combination with many other existing supervised machine learning programs available in or via Stata. \ddml has been tested with \texttt{lassopack} \citep{Ahrens2019}, \texttt{rforest} \citep{Schonlau2020}, \texttt{svmachines} \citep{Guenther2018}, and \texttt{parsnip} \citep{Huntington2021}. Indeed, the requirements for compatibility with \ddml are minimal: Any \texttt{eclass} program with the Stata-typical ``\texttt{reg\,\,y\,\,x}'' syntax, support for \texttt{if} conditions and post-estimation \texttt{predict} is compatible with \ddml. 

\item \ddml provides flexible multi-line syntax and short one-line syntax. The multi-line syntax offers a wide range of options, guides the user through the DDML algorithm step-by-step, and includes auxiliary programs for storing, loading and displaying additional information. We also provide a complementary one-line version called \qddml (`quick' \ddml), which uses a similar syntax as \pdslasso and \texttt{ivlasso} \citep{Ahrens2018}. 
\end{enumerate}

The article proceeds as follows. Section~\ref{sec:ddml_confounding} outlines DDML for the Partially Linear and Interactive Models under conditional unconfoundedness assumptions. Section~\ref{sec:ddml_iv} outlines DDML for Instrumental Variables (IV) models. Section~\ref{sec:stacking} discusses how stacking can be combined with DDML and provides evidence from Monte Carlo simulations illustrating the advantages of DDML with stacking. Section~\ref{sec:program} explains the features, syntax and options of the program. Section~\ref{sec:application} demonstrates the program's usage with two applications. 

\section{DDML with Conditional Unconfoundedness}\label{sec:ddml_confounding}

This section discusses DDML for the Partially Linear Model and the Interactive Model in turn. \revised{The exposition follows \citet{Chernozhukov2018}.} Both models are special cases of the general causal model \begin{align}\label{eq:all_causes_1}
    Y = f_0(D,\bm{X},U),
\end{align}
where $f_0$ is a structural function, $Y$ is the outcome, $D$ is the variable of interest, $\bm{X}$ are observed covariates, and $U$ are all unobserved determinants of $Y$ (i.e., other than $D$ and $\bm{X}$).\footnote{Since in \eqref{eq:all_causes_1}, $(D, \bm{X}, U)$ jointly determine $Y$, the model is also dubbed the ``all causes model'' (see, e.g., \citealp{heckman2007econometric}). Note that the model can equivalently be put into potential outcome notation with potential outcomes defined as $Y(d)\equiv f_0(d, \bm{X}, U)$.} The key difference between the Partially Linear Model and the Interactive Model is their position in the trade-off between functional form restrictions on $f_0$ and restrictions on the joint distribution of observables $(D,\bm{X})$ and unobservables $U$. For both models, we highlight key parameters of interest, state sufficient identifying assumptions, and outline the corresponding DDML estimator. A random sample $\{(Y_i, D_i, \bm{X}_i)\}_{i=1}^n$ from $(Y, D, \bm{X})$ is considered throughout.

\subsection{The Partially Linear Model (\texttt{partial})}\label{sec:plm}
The Partially Linear Model imposes the estimation model
\begin{align}\label{eq:plm}
\begin{aligned}
    Y & = \theta_0 D + g_0(\bm{X}) + U
    \end{aligned}
\end{align}
where $\theta_0$ is a fixed unknown parameter. The key feature of the model is that the controls $\bm{X}$ enter through the unknown and potentially nonlinear function $g_0$. Note that $D$ is not restricted to be binary and may be discrete, continuous or mixed. For simplicity, we assume that $D$ is a scalar, although \ddml allows for multiple treatment variables in the Partially Linear Model.

The parameter of interest is $\theta_0$, the causal effect of $D$ on $Y$.\footnote{The interpretation of $\theta_0$ can be generalized. For example, the results of \citet{angrist1999empirical} imply that in the general causal model \eqref{eq:all_causes_1}, $\theta_0$ is a positively weighted average of causal effects (e.g., conditional average treatment effects) under stronger identifying assumptions. The basic structure can also be used to obtain valid inference on objects of interest, such as projection coefficients, in the presence of high-dimensional data or nonparametric estimation without requiring a causal interpretation.} The key identifying assumption is given in Assumption \ref{assumption:plm_orthogonal}.\footnote{Discussions of Partially Linear Model typically show identification under the stronger assumption that $E[U\vert D,\bm{X}]=0$. We differentiate here to highlight differences between the Partially Linear Model and Interactive Model.} 
\begin{assumption}[Conditional Orthogonality]\label{assumption:plm_orthogonal}
$E[Cov(U, D\vert \bm{X})]= 0$ .
\end{assumption}

To show identification of $\theta_0$, consider the score \begin{align}\label{eq:plm_score}
    \psi(\bm{W}; \theta, m,\ell) = \Big(Y - \ell(\bm{X}) - \theta\big(D - m(\bm{X})\big)\Big)\big(D - m(\bm{X})\big),
\end{align}
where $\bm{W} \equiv (Y,D,\bm{X})$, and $\ell$ and $m$ are nuisance functions. Letting $m_0(\bm{X})\equiv E[D\vert \bm{X}] $ and $\ell_0(\bm{X})\equiv E[Y\vert \bm{X}]$, note that  
\begin{align*}
    E\left[\psi(\bm{W}; \theta_0, m_0, \ell_0)\right] = 0
\end{align*}
by Assumption \ref{assumption:plm_orthogonal}. When in addition $E[Var(D\vert \bm{X})] \neq 0$, we get 
\begin{align}\label{eq:plm_alpha0}
    \theta_0 = \frac{E\left[\big(Y - \ell_0(\bm{X})\big)\big(D - m_0(\bm{X})\big)\right]}{E\left[(D - m_0(\bm{X}))^2\right]}.
\end{align}

Equation \eqref{eq:plm_alpha0} is constructive in that it motivates estimation of $\theta_0$ via a simple two-step procedure: First, estimate the conditional expectation of $Y$ given $\bm{X}$ (i.e., $\ell_0$) and of $D$ given $\bm{X}$ (i.e., $m_0$) using appropriate nonparametric estimators (e.g., machine learners). Second, residualize $Y$ and $D$ by subtracting their respective conditional expectation function (CEF) estimates, and regress the resulting CEF residuals of $Y$ on the CEF residuals of $D$. This approach is fruitful when the estimation error of the first step does not propagate excessively to the second step. DDML leverages two key ingredients to control the impact of the first step estimation error on the second step estimate: 1) second step estimation based on Neyman orthogonal scores and 2) cross-fitting. As shown in \citet{Chernozhukov2018}, this combination facilitates the use of any nonparametric estimator that converges sufficiently quickly in the first and potentially opens the door for the use of many machine learners. 

Neyman orthogonality refers to a property of score functions $\psi$ that ensures local robustness to estimation errors in the first step. Formally, it requires that the Gateaux derivative with respect to the nuisance functions evaluated at the true values is mean-zero. In the context of the partially linear model, this condition is satisfied for the moment condition \eqref{eq:plm_score}:
\begin{align*}
    \partial_r \left\{E\left[\psi(W; \theta_0, m_0 + r(m - m_0),\ell_0 + r(\ell - \ell_0))\right]\right\}\vert_{r=0} = 0,
\end{align*} 
where the derivative is with respect to the scalar $r$ and evaluated at $r=0$. Heuristically, we can see that this condition alleviates the impact of noisy estimation of nuisance functions as local deviations of the nuisance functions away from their true values leave the moment condition unchanged. We refer to \citet{Chernozhukov2018} for a detailed discussion but highlight that all score functions discussed in this article are Neyman orthogonal.

Cross-fitting ensures independence between the estimation error from the first step and the regression residual in the second step. To implement cross-fitting, we randomly split the sample into $K$ evenly-sized folds, denoted as $I_1,\ldots, I_K$. For each fold $k$, the conditional expectations $\ell_0$ and $m_0$ are estimated using only observations \textit{not} in the $k$th fold -- i.e., in $I^c_k\equiv I \setminus I_k$ -- resulting in $\hat{\ell}_{I^c_{k}}$ and $\hat{m}_{I^c_{k}}$, respectively, where the subscript ${I^c_{k}}$ indicates the subsample used for estimation. The out-of-sample predictions for an observation $i$ in the $k$th fold are then computed via $\hat{\ell}_{I^c_{k}}(\bm{X}_i)$ and $\hat{m}_{I^c_{k}}(\bm{X}_i)$. Repeating this procedure for all $K$ folds then allows for computation of the DDML estimator for $\theta_0$:\begin{align}\label{eq:plm_estimator}
        \hat{\theta}_n = \frac{\frac{1}{n}\sum_{i=1}^n \big(Y_i-\hat{\ell}_{I^c_{k_i}}(\bm{X}_i)\big)\big(D_i-\hat{m}_{I^c_{k_i}}(\bm{X}_i)\big)}{\frac{1}{n}\sum_{i=i}^n \big(D_i-\hat{m}_{I^c_{k_i}}(\bm{X}_i)\big)^2},
\end{align}
where $k_i$ denotes the fold of the $i$th observation.\footnote{We here omit the constant from the estimation stage. Since the residualized outcome and treatment may not be exactly mean-zero in finite samples, \ddml includes the constant by default in the estimation stage of partially linear models.}

We summarize the DDML algorithm for the Partially Linear Model in Algorithm~1:\footnote{Algorithm~1 corresponds to the `DML2' algorithm in \citet{Chernozhukov2018}. \citet[Remark~3.1]{Chernozhukov2018} recommend `DML2' over the alternative `DML1' algorithm, which fits the final estimator by fold.} 

\noindent\begin{minipage}{\linewidth}
\begin{sttech}[Algorithm 1. DDML for the Partially Linear Model.]
Split the sample $\{(Y_i, D_i, \bm{X}_i)\}_{i=1}^n$ randomly in $K$ folds of approximately equal size. Denote $I_k$ the set of observations included in fold $k$ and $I_k^c$ its complement.
\begin{enumerate}[nosep] 
    \item For each $k\in\{1,\ldots, K\}$:
    \begin{enumerate}
            \item Fit a CEF estimator to the sub-sample $I_k^c$ using $Y_i$ as the outcome and $\bm{X}_i$ as predictors. Obtain the out-of-sample predicted values $\hat{\ell}_{I^c_k}(\bm{X}_i)$ for $i\in I_k$.
            \item Fit a CEF estimator to the sub-sample $I_k^c$ using $D_i$ as the outcome and $\bm{X}_i$ as predictors. Obtain the out-of-sample predicted values $\hat{m}_{I^c_k}(\bm{X}_i)$ for $i\in I_k$.
    \end{enumerate}
    \item Compute \eqref{eq:plm_estimator}.
\end{enumerate}
\end{sttech}
\end{minipage}

\citet{Chernozhukov2018} give conditions on the joint distribution of the data, in particular on $g_0$ and $m_0$, and properties of the nonparametric estimators used for CEF estimation, such that $\hat{\theta}_n$ is consistent and asymptotically normal. Standard errors are equivalent to the conventional linear regression standard errors of $Y_i-\hat{\ell}_{I^c_{k_i}}(\bm{X}_i)$ on $D_i-\hat{m}_{I^c_{k_i}}(\bm{X}_i)$. \texttt{ddml} computes the DDML estimator for the Partially Linear Model using Stata's \texttt{regress}. All standard errors available for linear regression in Stata are also available in \ddml, including different heteroskedasticity and cluster-robust standard errors.\footnote{See \texttt{help regress\#\#vcetype} for available options.}

\paragraph{Remark 1: Number of folds.}
The number of cross-fitting folds is a necessary tuning choice. Theoretically, any finite value is admissable. \citet[][Remark~3.1]{Chernozhukov2018} report that four or five folds perform better than only using $K=2$. Based on our simulation experience, we find that more folds tends to lead to better performance as more data is used for estimation of conditional expectation functions, especially when the sample size is small. We believe that more work on setting the number of folds would be useful, but believe that setting $K = 5$ provides is likely a good baseline in many settings.

\paragraph{Remark 2: Cross-fitting repetitions.}
DDML relies on randomly splitting the sample into $K$ folds. We recommend running the cross-fitting procedure more than once using different random folds to assess randomness introduced via the sample splitting. \ddml facilitates this using the \stopt{rep}{integer} options, which automatically estimates the same model multiple times and combines the resulting estimates to obtain the final estimate. By default, \ddml reports the median over cross-fitting repetitions. \ddml also supports the average of estimates. Specifically, let $\hat\theta_n^{(r)}$ denote the DDML estimate from the $r$th cross-fit repetition and $\hat{s}_n^{(r)}$ its associated standard error estimate with $r=1,\ldots,R$. The aggregate median point estimate and associated standard error are defined as
\[ \breve{\hat{\theta}}_n = {\textrm{median}}{\left(\left(\hat\theta_n^{(r)}\right)_{r=1}^R\right)} \quad\textrm{and}\quad \breve{\hat{s}}_n = \sqrt{\textrm{median}{\left(\left((\hat{s}_n^{(r)})^2 + (\hat\theta_n^{(r)} - \breve{\hat{\theta}}_n)^2\right)_{r=1}^R\right)}}. \]
The aggregate mean point estimates and associated standard error are calculated as
\[ \bar{\hat{\theta}}_n = \frac{1}{R}\sum_{r=1}^R \hat\theta_n^{(r)}  \quad\textrm{and}\quad \bar{\hat{s}}_n = \sqrt{\textrm{hmean}{\left(\left((\hat{s}_n^{(r)})^2 + (\hat\theta_n^{(r)} - \bar{\hat{\theta}}_n)^2\right)_{r=1}^R\right)}}, \]
where $\textrm{hmean}{()}$ is the harmonic mean.\footnote{The harmonic mean of $x_1,\ldots,x_n$ is defined as $\textrm{hmean}{(x_1,\ldots,x_n)}=n\left(\sum_{i=1}^n \frac{1}{x_i} \right)^{-1}$. We use the harmonic mean as it is less sensitive to outlier values.}

\paragraph{Remark 3: Cluster-dependence and folds.}
Under cluster-dependence, we recommend randomly assigning folds by cluster; see \stopt{fcluster}{varname}. 

\subsection{The Interactive Model (\texttt{interactive})}\label{sec:interactive}

The Interactive Model is given by 
\begin{align}\label{eq:interactive}
\begin{aligned}
    Y &= g_0(D,\bm{X}) + U 
\end{aligned}
\end{align}
where $D$ takes values in $\{0,1\}$. The key deviations from the Partially Linear Model are that $D$ must be a scalar binary variable and that $D$ is not required to be additively separable from the controls $\bm{X}$. In this setting, the parameters of interest we consider are\begin{align*}
    \theta^{\textrm{ATE}}_0 &\equiv E[g_0(1, \bm{X}) - g_0(0,\bm{X})]\\
    \theta^{\textrm{ATET}}_0 & \equiv E[g_0(1, \bm{X}) - g_0(0,\bm{X})\vert D = 1],
\end{align*}
which correspond to the average treatment effect (ATE) and average treatment effect on the treated (ATET), respectively.

Assumptions \ref{assumption:interactive_cmi} and \ref{assumption:interactive_overlap} below are sufficient for identification of the ATE and ATET. Note the conditional mean independence condition stated here is stronger than the conditional orthogonality assumption sufficient for identification of $\theta_0$ in the Partially Linear Model. 
\begin{assumption}[Conditional Mean Independence]\label{assumption:interactive_cmi}
$E[U\vert D, \bm{X}]=0$.
\end{assumption}
\begin{assumption}[Overlap]\label{assumption:interactive_overlap}
$\Pr(D=1 \vert \bm{X})\in (0,1)$ with probability 1.
\end{assumption}

Under assumptions \ref{assumption:interactive_cmi} and \ref{assumption:interactive_overlap}, we have\begin{align*}
    E[Y\vert D, \bm{X}] = E[g_0(D,\bm{X})\vert D, \bm{X}] + E[U\vert D, \bm{X}] = g_0(D,\bm{X}),
\end{align*}
so that identification of the ATE and ATET immediately follows from their definition.\footnote{In the defined Interactive Model under Assumption \ref{assumption:interactive_cmi}, the heterogeneity in treatment effects that the ATE and ATET average over is fully observed since $U$ is additively separable. Under stronger identifying assumptions, the DDML ATE and ATET estimators outlined here also apply to the ATE and ATET in the general causal model \eqref{eq:all_causes_1} that average over both observed and unobserved heterogeneity. See, e.g., \cite{Belloni2017}.} 

In contrast to Section \ref{sec:plm}, second-step estimators are not directly based on the moment conditions used for identification. Additional care is needed to ensure local robustness to first-stage estimation errors (i.e., Neyman orthogonality). In particular, the Neyman orthogonal score for the ATE that \citet{Chernozhukov2018} consider is the efficient influence function of \citet{Hahn1998}
\begin{align*}
    \psi^{\textrm{ATE}}(\bm{W};\theta, g, m) =  \frac{D(Y-g(1, \bm{X}))}{m(\bm{X})} {-} \frac{(1-D)(Y - g(0, \bm{X}))}{1 - m(\bm{X})} {+} g(1, \bm{X}) {-} g(0, \bm{X}) {-} \theta,
\end{align*}
where $\bm{W}\equiv (Y, D, \bm{X})$. Similarly for the ATET, 
\begin{align*}
    \psi^{\textrm{ATET}}(\bm{W};\theta, g, m, p) =  \frac{D(Y - g(0,\bm{X}))}{p} - \frac{m(\bm{X})(1-D)(Y-g(0,\bm{X}))}{p(1-m(\bm{X}))} -  \frac{D\theta}{p}. 
\end{align*}
Importantly, for $g_0(D,\bm{X})\equiv E[Y\vert D, \bm{X}]$, $m_0(\bm{X})\equiv E[D\vert \bm{X}]$, and $p_0 \equiv E[D]$, Assumptions \ref{assumption:interactive_cmi} and \ref{assumption:interactive_overlap} imply 
\begin{align*}E[\psi^{\textrm{ATE}}(\bm{W};\theta_0^{\textrm{ATE}}, g_0, m_0)] &= 0 \\  E[\psi^{\textrm{ATET}}(\bm{W};\theta_0^{\textrm{ATET}}, g_0, m_0, p_0)] &= 0;\end{align*}
and we also have that the Gateaux derivative of each condition with respect to the nuisance parameters $(g_0,m_0,p_0)$ is zero.

As before, the DDML estimators for the ATE and ATET leverage cross-fitting. The DDML estimators of the ATE and ATET based on $\psi^{\textrm{ATE}}$ and $\psi^{\textrm{ATET}}$ are 
\revised{\begin{align}
\begin{split}\label{eq:ddml_ate}
   \hat{\theta}^{\textrm{ATE}}_n &=\frac{1}{n}\sum_{i=1}^n\Bigg(\frac{D_i(Y_i-\hat{g}_{I^c_{k_i}}(1,\bm{X}_i))}{\hat{m}_{I^c_{k_i}}(\bm{X}_i)} - \frac{(1-D_i)(Y_i-\hat{g}_{I^c_{k_i}}(0,\bm{X}_i))}{1-\hat{m}_{I^c_{k_i}}(\bm{X}_i)} \\ &\hspace{.5\linewidth} +\hat{g}_{I^c_{k_i}}(1,\bm{X}_i)-\hat{g}_{I^c_{k_i}}(0,\bm{X}_i)\Bigg),
   \end{split}
   \\
   \begin{split}
      \hat{\theta}_n^{\textrm{ATET}} &=\frac{1}{n}\sum_{i=1}^n\left(\frac{D_i(Y_i-\hat{g}_{I^c_{k_i}}(0,\bm{X}_i))}{\hat{p}_{I^c_{k_i}}} - \frac{\hat{m}_{I^c_{k_i}}(\bm{X}_i)(1-D_i)(Y_i-\hat{g}_{I^c_{k_i}}(0,\bm{X}_i))}{\hat{p}_{I^c_{k_i}}(1-\hat{m}_{I^c_{k_i}}(\bm{X}_i))}\right)  ,\label{eq:ddml_atet}
      \end{split}
\end{align}}
where $\hat{g}_{I^c_{k}}$ and $\hat{m}_{I^c_{k}}$ are cross-fitted estimators for $g_0$ and $m_0$ as defined in Section \ref{sec:plm}. Since $D$ is binary, the cross-fitted values $\hat{g}_{I^c_{k}}(1,\bm{X})$ and $\hat{g}_{I^c_{k}}(0,\bm{X})$ are computed by only using treated and untreated observations, respectively.
\revised{$\hat{p}_{I^c_{k}}$ is a cross-fitted estimator of the unconditional treatment probability.} 

\ddml supports heteroskedasticity and cluster-robust standard errors for $\hat{\theta}_n^{\textrm{ATE}}$ and $\hat{\theta}_n^{\textrm{ATET}}$. 
The algorithm for estimating the ATE and ATET are conceptually similar to Algorithm~1. We delegate the detailed outline to Algorithm~A.1 in the Appendix. Mean and median aggregation over cross-fitting repetitions are implemented as outlined in Remark~2.

\section{DDML with Instrumental Variables}\label{sec:ddml_iv}
This section outlines the Partially Linear IV Model, the Flexible Partially Linear IV Model, and the Interactive IV Model. \revised{The discussion is again based on \citet{Chernozhukov2018}.}  As in the previous section, each model is a special case of the general causal model \eqref{eq:all_causes_1}. The discussion in this section differs from the preceding section in that identifying assumptions leverage instrumental variables $\bm{Z}$. The two partially linear IV models assume strong additive separability as in \eqref{eq:plm}, while the Interactive IV Model allows for arbitrary interactions between the treatment $D$ and the controls $\bm{X}$ as in \eqref{eq:interactive}. The Flexible Partially Linear IV Model allows for approximation of optimal instruments\footnote{We only accommodate approximation of optimal instruments under homoskedasticity. The instruments are valid more generally but are not optimal under heteroskedasticity. Obtaining optimal instruments under heteroskedasticity would require estimating conditional variance functions.} as in \citet{Belloni2012} and \citet{Chernozhukov2015a}, but relies on a stronger independence assumption than the Partially Linear IV Model. Throughout this discussion, we consider a random sample $\{(Y_i, D_i, \bm{X}_i,\bm{Z}_i)\}_{i=1}^n$ from $(Y, D, \bm{X}, \bm{Z})$.

\subsection{Partially Linear IV Model (\texttt{iv})}\label{sec:pliv}
The Partially Linear IV Model considers the same functional form restriction on the causal model as the Partially Linear Model in Section \ref{sec:plm}. Specifically, the Partially Linear IV Model maintains
\begin{align*}
\begin{aligned}
        Y&= \theta_0 D  + g_0(\bm{X}) + U, \\
\end{aligned}
\end{align*}
where $\theta_0$ is the unknown parameter of interest.\footnote{As in Section \ref{sec:plm}, the interpretation of $\theta_0$ can be generalized under stronger identifying assumptions. See \citet{angrist2000interpretation}.} 

The key deviation from the Partially Linear Model is that the identifying assumptions leverage instrumental variables $Z$, instead of directly restricting the dependence of $D$ and $U$. For ease of exposition, we focus on scalar-valued instruments in this section but we emphasize that \ddml for Partially Linear IV supports multiple instrumental variables and multiple treatment variables. 

Assumptions \ref{assumption:pliv_orthogonal} and \ref{assumption:pliv_relevance} below are sufficient orthogonality and relevance conditions, respectively, for identification of $\theta_0$.
\begin{assumption}[Conditional IV Orthogonality]\label{assumption:pliv_orthogonal}
$E[Cov(U,Z\vert \bm{X})]= 0$.
\end{assumption}
\begin{assumption}[Conditional Linear IV Relevance]\label{assumption:pliv_relevance}
$E[Cov(D,Z\vert \bm{X})]\neq 0$.
\end{assumption}

To show identification, consider the score function \begin{align*}
    \psi(\bm{W}; \theta, \ell,m,r) = \Big(Y-\ell(\bm{X}) - \theta(D - m(\bm{X}))\Big)\Big(Z - r(\bm{X})\Big),
\end{align*}
where $\bm{W}\equiv (Y, D, \bm{X}, Z)$. Note that for $\ell_0(\bm{X}) \equiv E[Y\vert \bm{X}]$, $m_0(\bm{X})\equiv E[D\vert \bm{X}]$, and $r_0(\bm{X})\equiv E[Z\vert \bm{X}]$, Assumption \ref{assumption:pliv_orthogonal} implies $E[\psi(\bm{W}; \theta_0, \ell_0,m_0,r_0)]=0.$ We will also have that the Gateux derivative of $E[\psi(\bm{W}; \theta_0, \ell_0,m_0,r_0)]$ with respect to the nuisance functions $(\ell_0,m_0,r_0)$ will be zero. Rewriting $E[\psi(\bm{W}; \theta_0, \ell_0,m_0,r_0)]=0$ then results in a Wald expression given by \begin{align}\label{eq:pliv_alpha0}
    \theta_0 = \frac{E\left[\big(Y-\ell_0(\bm{X})\big)\big(Z - r_0(\bm{X})\big)\right]}{E\left[\big(D - m_0(\bm{X})\big)\big(Z - r_0(\bm{X})\big)\right]},
\end{align}
where Assumption \ref{assumption:pliv_relevance} is used to ensure a non-zero denominator. 

The DDML estimator based on Equation \eqref{eq:pliv_alpha0} is given by \begin{align}\label{eq:pliv_ddml_estimator}
    \hat{\theta}_{n} = \frac{\frac{1}{n}\sum_{i=1}^n \left(Y_i-\hat{\ell}_{I^c_{k_i}}(\bm{X}_i)\right)\left(Z_i-\hat{r}_{I^c_{k_i}}(\bm{X}_i)\right)}{\frac{1}{n}\sum_{i=i}^n \left(D_i-\hat{m}_{I^c_{k_i}}(\bm{X}_i)\right)\left(Z_i-\hat{r}_{I^c_{k_i}}(\bm{X}_i)\right)},
\end{align}
where $\hat{\ell}_{I^c_{k}}$, $\hat{m}_{I^c_{k}}$, and $\hat{r}_{I^c_{k}}$ are appropriate cross-fitted CEF estimators.

Standard errors corresponding to $\hat{\theta}_{n}$ are equivalent to the IV standard errors where $Y_i-\hat{\ell}_{I^c_{k_i}}(\bm{X}_i)$ is the outcome, $D_i-\hat{m}_{I^c_{k_i}}(\bm{X}_i)$ is the endogenous variable, and $Z_i-\hat{r}_{I^c_{k_i}}(\bm{X}_i)$ is the instrument. \ddml supports conventional standard errors available for linear instrumental variable regression in Stata, including heteroskedasticity and cluster-robust standard errors. Mean and median aggregation over cross-fitting repetitions are implemented as outlined in Remark~2. In the case where we have multiple instruments or endogenous regressors, we adjust the algorithm by residualizing each instrument and endogenous variable as above and applying two-stage least squares with the residualized outcome, endogenous variables, and instruments.

\subsection{Flexible Partially Linear IV Model (\texttt{fiv})}\label{sec:hdiv}
The Flexible Partially Linear IV Model considers the same parameter of interest as the Partially Linear IV Model. The key difference here is that identification is based on a stronger independence assumption which allows for approximating optimal instruments using nonparametric estimation, including machine learning, akin to \citet{Belloni2012} and \citet{Chernozhukov2015a}.
In particular, the Flexible Partially Linear IV Model leverages a conditional mean independence assumption rather than an orthogonality assumption as in Section \ref{sec:pliv}. As in Section \ref{sec:pliv}, we state everything in the case of a scalar $D$. 

\begin{assumption}[Conditional IV Mean Independence]\label{assumption:hdiv_iv_meanindependence}
$E[U\vert \bm{Z},\bm{X}] = 0$.
\end{assumption}
Assumption \ref{assumption:hdiv_iv_meanindependence} implies that for any function $\tilde{p}(\bm{Z},\bm{X})$, it holds that  \begin{align}\label{eq:hdiv_moment}
    E\left[\Big(Y - \ell_0(X) - \theta\big(D - m_0(\bm{X})\big)\Big)\Big(\tilde{p}(\bm{Z},\bm{X}) - E\left[\tilde{p}(\bm{Z},\bm{X})\vert \bm{X}\right]\Big)\right] = 0,
\end{align}
where $\ell_0(\bm{X})=E[Y\vert \bm{X}]$ and $m_0(\bm{X})=E[D\vert \bm{X}]$. Identification based on \eqref{eq:hdiv_moment} requires that there exists some function $\tilde{p}$ such that \begin{align}\label{eq:hdiv_ptld_relevance}
    E\left[Cov(D, \tilde{p}(\bm{Z},\bm{X})\vert \bm{X})\right] \neq 0.
\end{align}
A sufficient assumption is that $D$ and $\bm{Z}$ are not mean independent conditional on $\bm{X}$. This condition allows setting $\tilde{p}(\bm{Z},\bm{X}) = E[D\vert \bm{Z},\bm{X}]$ which will then satisfy \eqref{eq:hdiv_ptld_relevance}.\footnote{The choice $\tilde{p}(\bm{Z},\bm{X}) = E[D\vert \bm{Z},\bm{X}]$ results in the optimal instrument, in the sense of semi-parametric efficiency, under homoskedasticity.} Assumption \ref{assumption:hdiv_relevance} is a consequence of this non-mean independence.
\begin{assumption}[Conditional IV Relevance]\label{assumption:hdiv_relevance}
$E[Var(E[D\vert \bm{Z},\bm{X}]\vert \bm{X})] \neq 0.$ 
\end{assumption}

Consider now the score function \begin{align*}
    \psi(\bm{W}; \theta, \ell, m, p) = \Big(Y-\ell(\bm{X}) - \theta(D - m(\bm{X}))\Big)\Big(p(\bm{Z}, \bm{X}) - m(\bm{X})\Big),
\end{align*}
where $\bm{W}\equiv (Y, D, \bm{X}, \bm{Z})$. Note that for  $\ell_0(\bm{X}) \equiv E[Y\vert \bm{X}]$, $m_0(\bm{X})\equiv E[D\vert \bm{X}]$, and $p_0(\bm{Z},\bm{X}) \equiv E[D\vert \bm{Z},\bm{X}]$, Assumption \ref{assumption:hdiv_iv_meanindependence} and the law of iterated expectations imply $E[\psi(\bm{W}; \theta_0, \ell_0, m_0, p_0)]=0$ and the Gateaux differentiability condition holds. Rewriting then results in a Wald expression given by \begin{align}\label{eq:hdiv_alpha0}
    \theta_0 = \frac{E\left[\big(Y-\ell_0(\bm{X})\big)\big(p_0(\bm{Z}, \bm{X}) - m_0(\bm{X})\big)\right]}{E[\big(D-m_0(\bm{X})\big)\big(p_0(\bm{Z}, \bm{X}) - m_0(\bm{X})\big)]},
\end{align}
where Assumption \ref{assumption:hdiv_relevance} ensures a non-zero denominator.

The DDML estimator based on the moment solution \eqref{eq:hdiv_alpha0} is given by \begin{align}\label{eq:hdiv_ddml_estimator}
        \hat{\theta}_{n} = \frac{\frac{1}{n}\sum_{i=1}^n \big(Y_i-\hat{\ell}_{I^c_{k_i}}(\bm{X}_i)\big)\big(\hat{p}_{I^c_{k_i}}(\bm{Z}_i, \bm{X}_i)-\hat{m}_{I^c_{k_i}}(\bm{X}_i)\big)}{\frac{1}{n}\sum_{i=i}^n \big(D_i-\hat{m}_{I^c_{k_i}}(\bm{X}_i)\big)\big(\hat{p}_{I^c_{k_i}}(\bm{Z}_i, \bm{X}_i)-\hat{m}_{I^c_{k_i}}(\bm{X}_i)\big)},
\end{align}
where $\hat{\ell}_{I^c_{k}}$, $\hat{m}_{I^c_{k}}$, and $\hat{p}_{I^c_{k}}$ are appropriate cross-fitted CEF estimators.

In simulations, we find that the finite sample performance of the estimator in \eqref{eq:hdiv_ddml_estimator} improves when the law of iterated expectations applied to $E[p_0(\bm{Z},\bm{X})] = m_0(\bm{X})$ is explicitly approximately enforced in estimation. As a result, we propose an intermediate step to the previously considered two-step DDML algorithm: Rather than estimating the conditional expectation of $D$ given $\bm{X}$ directly, we estimate it by projecting first-step estimates of the conditional expectation of $p_0(\bm{Z},\bm{X})$ onto $\bm{X}$ instead. Algorithm~2 outlines the LIE-compliant DDML algorithm for computation of \eqref{eq:hdiv_ddml_estimator}. 

\noindent\begin{minipage}{\linewidth}
\begin{sttech}[Algorithm 2. LIE-compliant DDML for the Flexible Partially Linear IV Model.]
 Split the sample $\{(Y_i, D_i, \bm{X}_i, \bm{Z}_i)\}_{i=1}^n$ randomly in $K$ folds of approximately equal size. Denote $I_k$ the set of observations included in fold $k$ and $I_k^c$ its complement.
\begin{enumerate}[nosep]  
    \item For each $k\in\{1,\ldots, K\}$:
    \begin{enumerate}
            \item Fit a CEF estimator to the sub-sample $I_k^c$ using $Y_i$ as the outcome and $\bm{X}_i$ as predictors. Obtain the out-of-sample predicted values $\hat{\ell}_{I^c_k}(\bm{X}_i)$ for $i\in I_k$.
            \item Fit a CEF estimator to the sub-sample $I_k^c$ using $D_i$ as the outcome and $(\bm{Z}_i,\bm{X}_i)$ as predictors. Obtain the out-of-sample predicted values $\hat{p}_{I^c_k}(\bm{Z}_i,\bm{X}_i)$ for $i\in I_k$ and in-sample predicted values $\hat{p}_{I^c_k}(\bm{Z}_i,\bm{X}_i)$ for $i\in I^c_k$.
            \item Fit a CEF estimator to the sub-sample $I_k^c$ using the in-sample predicted values $\hat{p}_{I^c_k}(\bm{Z}_i,\bm{X}_i)$ as the outcome and $X_i$ as predictors. Obtain the out-of-sample predicted values $\hat{m}_{I^c_k}(\bm{X}_i)$ for $i\in I_k.$
    \end{enumerate}
    \item Compute \eqref{eq:hdiv_ddml_estimator}.
\end{enumerate}
\end{sttech}
\end{minipage}

Standard errors corresponding to $\hat{\theta}_{n}$ in  \eqref{eq:hdiv_ddml_estimator} are the same as in Section \ref{sec:pliv} where the instrument is now given by $\hat{p}_{I^c_{k_i}}(\bm{Z}_i, \bm{X}_i)-\hat{m}_{I^c_{k_i}}(\bm{X}_i)$. Mean and median aggregation over cross-fitting repetitions are as outlined in Remark~2. 

\subsection{Interactive IV Model (\texttt{interactiveiv})}\label{sec:interactiveiv}

The Interactive IV Model considers the same causal model as in Section \ref{sec:interactive}; specifically
\begin{align*}
\begin{aligned}
    Y &= g_0(D, \bm{X}) + U \\
\end{aligned}
\end{align*}
where $D$ takes values in $\{0,1\}$. The key difference from the Interactive Model is that this section considers identification via a binary instrument $Z$ \revised{representing assignment to treatment}.

The parameter of interest we target is\begin{align}\label{interactiveiv:fuzzy_late}
\begin{aligned}
   \theta_0 & =  E\left[g_0(1, \bm{X}) - g_0(0, \bm{X})\vert \: p_0(1,\bm{X}) > p_0(0,\bm{X})\right],
\end{aligned}
\end{align}
where $p_0(Z,\bm{X})\equiv \Pr(D=1\vert Z, \bm{X})$. Here, $\theta_0$ is a local average treatment effect (LATE). \revised{Note that in contrast to the LATE developed in \citet{imbens1994identification}, we follow the exposition in \citet{Chernozhukov2018} where ``local'' does not strictly refer to compliers but instead observations with a higher propensity score -- i.e., a higher probability of complying.}\footnote{\revised{Identification of the conventional complier-focused LATE is achieved under stronger conditional independence and monotonicity assumptions not introduced in this paper. Under these stronger assumptions, the DDML LATE estimator outlined here targets the conventionally considered LATE parameter.}}

Identification again leverages Assumptions \ref{assumption:hdiv_iv_meanindependence} and \ref{assumption:hdiv_relevance} made in the context of the Flexible Partially Linear IV Model. In addition, we assume that the propensity score is weakly monotone with probability one, and that the support of the instrument is independent of the controls. 
\begin{assumption}[Monotonicity]\label{assumption:iv_monotonicty}
$p_0(1, \bm{X})\geq p_0(0, \bm{X})$ with probability 1.
\end{assumption} 
\begin{assumption}[IV Overlap]\label{assumption:iv_overlap}
$\Pr(Z=1\vert \bm{X})\in(0,1)$ with probability 1.
\end{assumption}

Assumptions \ref{assumption:hdiv_iv_meanindependence}-\ref{assumption:iv_overlap} imply that \begin{align}\label{eq:interactiveiv_flate}
    \theta_0 = \frac{E\left[\ell_0(1, \bm{X})-\ell_0(0, \bm{X})\right]}{E\left[p_0(1, \bm{X})-p_0(0, \bm{X})\right]},
\end{align}
where $\ell_0(Z, \bm{X}) \equiv E[Y\vert Z, \bm{X}]$, verifying identification of the LATE $\theta_0$. Akin to Section \ref{eq:interactive}, however, estimators of $\theta_0$ should not directly be based on Equation \eqref{eq:interactiveiv_flate} because the estimating equations implicit in obtaining \eqref{eq:interactiveiv_flate} do not satisfy Neyman-orthogonality. Hence, a direct estimator of $\theta_0$ obtained by plugging nonparametric estimators in for nuisance functions in  \eqref{eq:interactiveiv_flate} will potentially be highly sensitive to the first step nonparametric estimation error. Rather, we base estimation on the Neyman orthogonal score function \begin{align*}
    \psi(\bm{W};\theta, \ell, p, r) &= \frac{Z(Y - \ell(1, \bm{X}))}{r(\bm{X})} - \frac{(1-Z)(Y-\ell(0, \bm{X}))}{1-r(\bm{X})} + \ell(1, \bm{X})-\ell(0, \bm{X}) \\
    &  + \left[\frac{Z(D - p(1,\bm{X}))}{r(\bm{X})} - \frac{(1-Z)(D-p(0,\bm{X}))}{1-r(\bm{X})} + p(1,\bm{X})-p(0,\bm{X})\right] \times \theta
\end{align*}
where $\bm{W} \equiv (Y, D, \bm{X}, Z)$ . Note that under Assumptions \ref{assumption:hdiv_iv_meanindependence}-\ref{assumption:iv_overlap} and for $\ell_0(Z,\bm{X}) \equiv E[Y\vert Z,\bm{X}]$, $p_0(Z,\bm{X}) \equiv E[D|Z,\bm{X}]$, and $r_0(\bm{X})\equiv E[Z\vert \bm{X}]$, we have $E[\psi(\bm{W};\theta_0, \ell_0, p_0, r_0)]=0$ and can verify that its Gateaux derivative with respect to the nuisance functions local to their true values is also zero.

The DDML estimator based on the orthogonal score $\psi$ is then \begin{align}
\begin{split}
    \hat{\theta}_n &= \\
    &\frac{\frac{1}{n}\sum_{i}\left(\frac{Z_i(Y_i-\hat{\ell}_{I^c_{k_i}}(1,\bm{X}_i))}{\hat{r}_{I^c_{k_i}}(\bm{X}_i)} - \frac{(1-Z_i)(Y_i-\hat{\ell}_{I^c_{k_i}}(0,\bm{X}_i))}{1-\hat{r}_{I^c_{k_i}}(\bm{X}_i)}+\hat{\ell}_{I^c_{k_i}}(1,\bm{X}_i)-\hat{\ell}_{I^c_{k_i}}(0,\bm{X}_i)\right)}{\frac{1}{n}\sum_{i}\left(\frac{Z_i(D_i-\hat{p}_{I^c_{k_i}}(1,\bm{X}_i))}{\hat{r}_{I^c_{k_i}}(\bm{X}_i)} - \frac{(1-Z_i)(D_i-\hat{p}_{I^c_{k_i}}(0,\bm{X}_i))}{1-\hat{r}_{I^c_{k_i}}(\bm{X}_i)}+\hat{p}_{I^c_{k_i}}(1,\bm{X}_i)-\hat{p}_{I^c_{k_i}}(0,\bm{X}_i)\right)}, \label{eq:ddml_late}
\end{split}
\end{align}
where $\hat{\ell}_{I^c_{k}}$, $\hat{p}_{I^c_{k}}$, and $\hat{r}_{I^c_{k}}$ are appropriate cross-fitted CEF estimators. Since $Z$ is binary, the cross-fitted values $\hat{\ell}_{I^c_{k}}(1,\bm{X})$ and $\hat{p}_{I^c_{k}}(1,\bm{X})$, as well as $\hat{\ell}_{I^c_{k}}(0,\bm{X})$ and $\hat{p}_{I^c_{k}}(0,\bm{X})$ \revised{are computed by using only assigned and unassigned observations, respectively.} 

\ddml supports heteroskedasticity and cluster-robust standard errors for $\hat{\theta}_n$. Mean and median aggregation over cross-fitting repetitions are implemented as outlined in Remark~2. 

\section{The choice of machine learner}\label{sec:stacking}
\citet{Chernozhukov2018} show that DDML estimators are asymptotically normal when used in combination with a general class of machine learners satisfying a relatively weak convergence rate requirement for estimating the CEFs. While asymptotic properties of common machine learners remain an highly active research area, recent advances provide convergence rates for special instances of many machine learners, including lasso \citep{bickel2009simultaneous, Belloni2012}, random forests \citep{wager2015adaptive, wager2018, Athey2019a}, neural networks \citep{schmidt2020nonparametric, farrell2021deep}, and boosting \citep{luo2016high}. It seems likely that many popular learners will fall under the umbrella of suitable learners as theoretical results are further developed. However, we note that currently known asymptotic properties do not cover a wide range of learners, such as very deep and wide neural networks and deep random forests, as they are currently implemented in practice.  



The relative robustness of DDML to the first-step learners leads to the question of which machine learner is the most appropriate for a given application. It is \emph{ex ante} rarely obvious which learner will perform best. Further, rather than restricting ourselves to one learner, we might want to combine several learners into one final learner. This is the idea behind stacking generalization, or simply ``stacking'', due to \citet{Wolpert1992} and \citet{Breiman1996a}. Stacking allows one to accommodate a diverse set of base learners with varying tuning and hyper-tuning parameters. It thus provide a convenient framework for combining and identifying suitable learners, thereby reducing the risk of misspecification. \revised{\citet{Ahrens2023_applied} introduce short-stacking which reduces the computational cost of pairing DDML and stacking drastically, as well as pooled stacking which enforces common weights across cross-fitting folds.}

\revised{We discuss stacking approaches to DDML estimation in Section~\ref{eq:stacking}.} Section~\ref{eq:stacking_mc} demonstrates the performance of DDML in combination with stacking \revised{approaches} using a simulation.

\subsection{\revised{DDML and stacking}}\label{eq:stacking}
Our discussion of stacking in the context of DDML focuses on the Partially Linear Model in \eqref{eq:plm}, but we highlight that DDML and stacking can be combined in the same way for all other models supported in \ddml. Suppose we consider $J$ machine learners, referred to as base learners, to estimate the CEFs $\ell_0(\bm{X}) \equiv E[Y|\bm{X}]$ and $m_0(\bm{X}) \equiv E[D|\bm{X}]$. The set of base learners could, for example, include cross-validated lasso and ridge with alternative sets of predictors, gradient boosted trees with varying tree depth and feed-forward neural nets with varying number of hidden layers and neurons. Generally, we recommend considering a relatively large and diverse set of base learners, and including some learners with alternative tuning parameters. 

We randomly split the sample into $K$ cross-fitting folds, denoted as $I_1,\ldots,I_K$. In each cross-fitting step $k$, we define the training sample as $I_k^c\equiv T_k$, comprising all observations excluding the cross-fitting hold-out fold $k$. This training sample is further divided into $V$ cross-validation folds, denoted as $T_{k,1},\ldots,T_{k,V}$. The stacking regressor fits a final learner to the training sample $T_k$ using the cross-validated predicted values of each base learner as inputs. A typical choice for the final learner is constrained least squares (CLS) which restricts the weights to be positive and sum to one. The stacking objective function for estimating $\ell_0(\bm{X})$ using the training sample $T_k$ is then defined as:
\begin{align}\label{eq:stacking_dfn}
\underset{w_{k,1},\ldots,w_{k,J}}{\min} \sum_{i\in T_k} \left( Y_i - \sum_{j=1}^J w_{k,j} \hat{\ell}^{(j)}_{T_{k,v(i)}^c}(\bm{X}_i)  \right)^2,   \qquad \textrm{s.t.}\ w_{k,j}\geq 0,\ \sum_{j=1}^J |w_{k,j}|=1,
\end{align}
where $w_{k,j}$ are referred to as stacking weights. We use $\hat{\ell}^{(j)}_{T_{k,v(i)}^c}(\bm{X}_i)$ to denote the cross-validated predicted value for observation $i$, which is obtained from fitting learner $j$ on the sub-sample $T_{k,v(i)}^c \equiv T_k \setminus T_{k,v(i)}$, i.e., the sub-sample excluding the fold $v(i)$ \revised{into which observation $i$ falls}. The stacking predicted values are obtained as $\sum_j \hat{w}_{k,j} \hat{\ell}_k^{(j)}(\bm{X}_i)$ where each learner $j$ is fit on the step-$k$ training sample~$T_k$. The objective function for estimating $m_0(\bm{X})$ is defined accordingly. 

CLS frequently performs well in practice and facilitates the interpretation of stacking as a weighted average of base learners \citep{Hastie2009}. It is, however, not the only sensible choice of combining base learners. For example, stacking could instead select the single learner with the lowest quadratic loss, i.e., by imposing the constraint $w_{k,j}\in\{0,1\}$ and $\sum_{k,j}w_{k,j}=1$. We refer to this choice as ``single best'' and include it in our simulation experiments. We implement stacking for DDML using \pystacked \citep{Ahrens2022}.

\paragraph{Pooled stacking.} \revised{A variant of stacking specific to DDML is \emph{pooled stacking}. 
Standard stacking fits the final learner $K$ separate times, once in each cross-fitting step, yielding $K$ separate sets of stacking weights $\hat{w}_{k,j}$ for the $J$ learners.
With DDML pooled stacking, we can impose the additional constraint in \eqref{eq:stacking_dfn} that the weights are the same across all cross-fit folds, $\hat{w}_{k,j}=\hat{w}_{j}, \forall \: k$.
By returning a single set of stacking weights, pooled stacking imposes an additional degree of regularization and facilitates interpretation but suffers from the same high computational cost as pairing DDML with (regular) stacking.}

\paragraph{Short-stacking.} Stacking \revised{and pooled stacking rely} on cross-validation. In the context of DDML we can also exploit the cross{\it -fitted} predicted values directly for stacking. That is, we can directly apply CLS to the cross-fitted predicted values for estimating $\ell_0(\bm{X})$ (and similarly $m_0(\bm{X})$):
\[  \underset{w_1,\ldots,w_J}{\min} \sum_{i=1}^n \left( Y_i - \sum_{j=1}^J w_j\hat{\ell}^{(j)}_{I_{k(i)}^c}(\bm{X}_i)   \right)^2,   \qquad \textrm{s.t.}\ w_j\geq 0,\ \sum_{j=1}^J |w_j|=1\]
We refer to this form of stacking that utilizes the cross-fitted predicted values as \emph{short-stacking} as it takes a short-cut. This is to contrast it with regular stacking which estimates the stacking weights for each cross-fitting fold $k$. \revised{The main advantage of short-stacking relative to standard stacking is the lower computational cost, because short-stacking does not require the fitting of the $j$ learners on sub-samples to obtain the cross-validated predicted values $\hat{\ell}^{(j)}_{T_{k,v(i)}^c}(\bm{X}_i)$ needed for standard stacking.
Furthermore, short-stacking (like pooled stacking) also produces a single set of weights for the entire sample, which facilitates interpretation and implies a higher degree of regularization. A potential disadvantage of short-stacking is that it is more susceptible to over-fitting issues since stacking weights and structural parameters are estimated using the same cross-fitted predicted values. We thus recommend only considering short-stacking in regular settings where the number of candidate learners is small relative to $N$ \citep[see also the discussion in][]{Ahrens2023_applied}.} Algorithm~A.4 in the Appendix summarizes the short-stacking algorithm for the Partially Linear Model.\footnote{While short-stacking can be applied in a similar fashion to other conditional expectations, a complication arises in the Flexible Partially Linear IV Model where the cross-fitted predicted values of $E[D|\bm{X}]$ depend on $E[D|\bm{X},\bm{Z}]$. We describe the algorithm that accounts for this in the Appendix; see Algorithm~A.5.}

\subsection{Monte Carlo simulation}\label{eq:stacking_mc}
To illustrate the advantages of DDML with stacking, we generate artificial data based on the Partially Linear Model
\begin{align}
Y_i &= \theta_0D_i + c_Yg(\bm{X}_i) + \sigma_Y(D_i,\bm{X}_i)\varepsilon_i \label{eq:mc1}\\
D_i &= c_Dg(\bm{X}_i) + \sigma_D(\bm{X}_i)u_i \label{eq:mc2}
\end{align}
where both $\varepsilon_i$ and $u_i$ are independently drawn from the standard normal distribution. We set the target parameter to $\theta_0=0.5$ and the sample size to either $n=100$ or $n=1000$. The controls $\bm{X}_i$ are drawn from the multivariate normal distribution with $N(0,\bm{\Sigma})$ where $\Sigma_{ij}=(0.5)^{|i-j|}$. The number of controls is set to $p=\dim(\bm{X}_i)=50$, except in DGP~5 where $p=7$.  The constants $c_Y$ and $c_D$ are chosen such that the $R^2$ in \eqref{eq:mc1} and \eqref{eq:mc2}
are approximately equal to 0.5. To induce heteroskedasticity, we set 
\[\sigma_D\left(\bm{X}_i\right)=\sqrt{\frac{\left(1+g(\bm{X}_i)\right)^2}{\frac{1}{n}\sum_i\left(1+g(\bm{X}_i)\right)^2}} \quad\text { and } \quad \sigma_Y\left(D_i, \bm{X}_i\right)=\sqrt{\frac{\left(1+\theta_0 D_i+g(\bm{X}_i)\right)^2}{\frac{1}{n}\sum_i\left(1+\theta_0 D_i+g(\bm{X}_i)\right)^2}}\] 
The nuisance function $g(\bm{X}_i)$ is generated using five exemplary DGPs, which cover linear and nonlinear processes with varying degrees of sparsity and varying number of observed covariates:
\[
\begin{array}{ll}
\textrm{DGP 1:}\quad g(\bm{X}_i) &= \displaystyle\sum\nolimits_j 0.9^jX_{ij} \\
\textrm{DGP 2:}\quad g(\bm{X}_i) &= X_{i1}X_{i2}{+}X_{i3}^2{+}X_{i4}X_{i5}{+}X_{i6}X_{i7}{+}X_{i8}X_{i9}{+}X_{i10}{+} X_{i11}^2{+}X_{i12} X_{i13} \phantom{\displaystyle\sum\nolimits_j} \\
\textrm{DGP 3:}\quad g(\bm{X}_i) &= \mathbbm{1}\{X_{i1}>0.3\}\mathbbm{1}\{X_{i2}>0\}\mathbbm{1}\{X_{i3}>-1\}  \phantom{\displaystyle\sum\nolimits_j} \\
\textrm{DGP 4:}\quad g(\bm{X}_i) &= X_{i1} + \sqrt{|X_{i2}|} + \textrm{sin}(X_{i3}) + 0.3X_{i4}X_{i5} + X_{i6} + 0.3X_{i7}^2  \phantom{\displaystyle\sum\nolimits_j} \\
\textrm{DGP 5:}\quad g(\bm{X}_i)&=\textnormal{same as DGP 4 with } p=7 \phantom{\displaystyle\sum\nolimits_j} \\
\end{array}\]
DGP~1 is a linear design involving many negligibly small parameters. While not exactly sparse, the design can be approximated well through a sparse representation. DGP~2 is linear in the parameters and exactly sparse, but includes interactions and second-order polynomials. DGPs~3-5 are also exactly sparse but involve complex nonlinear and interaction effects. DGP~4 and~5 are identical, except that DGP~5 does not add nuisance covariates that are unrelated to $Y$ and $D$.

We consider DDML with the following supervised machine learners for cross-fitting the CEFs:\footnote{All base learners have been implemented using \pystacked. We use the defaults of \pystacked for parameter values and settings not mentioned here.} 
\begin{enumerate}[noitemsep,topsep=0pt,itemsep=0pt,parsep=0pt,before=\vspace{0mm},after=\vspace{-0mm}]
    \item[1.-2.] Cross-validated lasso \& ridge with untransformed base controls 
    \item[3.-4.] Cross-validated lasso \& ridge with 5th-order polynomials of base controls but no interactions (referred to as `Poly 5')
    \item[5.-6.] Cross-validated lasso \& ridge with second-order polynomials and all first order interaction terms (referred to as `Poly 2 + Inter.')
    \item[7.] Random forests (RF) with low regularization: base controls, maximum tree depth of 10, 500 trees and approximately $\sqrt{p}$ features considered at each split 
    \item[8.] RF with medium regularization: same as 7., but with maximum tree depth of 6 
    \item[9.] RF with high regularization: same as 7., but with maximum tree depth of 2 
    \item[10.] Gradient boosted trees (GB) with low regularization: base controls, 1000 trees and a learning rate of 0.3. We enable early stopping which uses a 20\% validation sample to decide whether to stop the learning algorithm. Learning is terminated after 5 iterations with no meaningful improvement in the mean-squared loss of the validation sample.\footnote{We use a tolerance level of 0.01 to measure improvements.} 
    \item[11.] GB with medium regularization: same as 10., but with learning rate of 0.1  
    \item[12.] GB with high regularization: same as 10., but with learning rate of 0.01 
    \item[13.] Feed-forward neural net with base controls and two layers of size 20 
\end{enumerate}

\noindent We use the above set of learners as base learners for DDML with stacking approaches. Specifically, \revised{we estimate DDML using stacking, short-stacking and pooled stacking which we combine with CLS and the single-best learner}. We set the number of folds to $K=20$ if $n=100$, and $K=5$ if $n=1000$. \revised{That is, we adapt the number of folds $K$ to the total sample size $n$ to ensure that the CEF estimators are trained on sufficiently large training samples.}

For comparison, we report results for \revised{OLS and PDS-lasso with base controls}, PDS-Lasso with Poly 5, PDS-Lasso with Poly 2 + Interactions, and an oracle estimator using the full sample.\footnote{The PDS-Lasso estimators set tuning parameters using the default in \pdslasso.} The oracle estimator presumes knowledge of the function $g(\bm{X})$, and obtains estimates by regressing $Y$ on the two variables $D$ and $g(\bm{X})$. 

We report simulation median absolute bias (MAB) and coverage rates of 95\% confidence intervals (CR) for DGPs 1-3 in Table~\ref{tab:mc1}. We delegate results for DGPs~4 and~5, including a brief discussion, to Appendix~B. DDML estimators leveraging stacking \revised{approaches} perform favorably in comparison to individual base learners in terms of bias and coverage. The relative performance of stacking approaches seems to improve as the sample size increases, likely reflecting that the stacking weights are \revised{more precisely estimated in larger small samples}. For $n=1000$, the bias of stacking with CLS is at least as low as the bias of the best-performing individual learner under DGP~1-2, while only gradient boosting and neural net yield a lower bias than stacking under DGP~3. 

Results for coverage are similar with stacking-based estimates being comparable with the best performing feasible estimates and the oracle when $n = 1000.$ With $n = 100$, coverage of confidence intervals for stacking-based estimators are inferior to coverages for a small number of the individual learners but still competitive and superior than most learners. Looking across all results, we see that stacking provides robustness to potentially very bad performance that could be obtained from using a single poorly performing learner.

There are overall little performance differences among the \revised{six stacking estimators considered., suggesting that short-stacking has a substantial practical advantage due to its lower computational cost. \citet{Ahrens2023_applied} report that short-stacking reduces the compute time by a factor of $1/V$ where $V$ is the number of cross-validation folds.} There is some evidence that the single-best selector outperforms CLS in very small sample sizes in DGPs 2-3, but not in DGP 1 (and also not in DGPs 4-5, see Table~B.1). We suspect that the single-best selector works better in scenarios where there is one base learner that clearly dominates. 

The mean-squared prediction errors (MSPE) and the average stacking weights, which we report in Tables~B.2 and B.3 in the Appendix, provide further insights into how stacking with CLS functions. CLS assigns large stacking weights to base learners with a low MSPE, which in turn are associated with a low bias. Importantly, stacking assigns zero or close-to-zero weights to poorly specified base learners such as the highly regularized random forest, which in all three DGPs ranks among the individual learners with highest MSPE and highest bias. The robustness to misspecified and ill-chosen machine learners, which could lead to misleading inference, is indeed one of our main motivations for advocating stacking approaches to DDML. 

DDML with stacking approaches also compares favorably to conventional full-sample estimators. In the relatively simple linear DGP~1, DDML with stacking performs similarly to OLS and the infeasible oracle estimator---both in terms of bias and coverage---for $n=100$ and $n=1000$. In the more challenging DGPs 2 and 3, the bias of DDML with stacking is substantially lower than the biases of OLS and the PDS-Lasso estimators. While the bias and size distortions of DDML with stacking are still considerable in comparison to the infeasible oracle for $n=100$, they are close to the oracle for $n=1000$. The results overall highlight the flexibility of DDML with stacking to flexibly approximate a wide range of DGPs provided a diverse set of base learners is chosen.

\begin{sidewaystable} \revisedcolor
    \centering\scriptsize\singlespacing
\begin{threeparttable}
    \caption{Bias and Coverage Rates in the Linear and Nonlinear DGPs}
    \label{tab:mc1}
        \begin{tabular}{rlcccccccccccccc}
    \toprule
    \midrule
          &       & \multicolumn{4}{c}{DGP 1} & & \multicolumn{4}{c}{DGP 2} & & \multicolumn{4}{c}{DGP 3} \\
    \cline{3-6} \cline{8-11} \cline{13-16} \\[-6pt]
     & & \multicolumn{2}{c}{$n=100$}    &\multicolumn{2}{c}{$n=1000$}    &&  \multicolumn{2}{c}{$n=100$}    &\multicolumn{2}{c}{$n=1000$}     &&  \multicolumn{2}{c}{$n=100$}    &\multicolumn{2}{c}{$n=1000$}       \\
    && MAB & CR   &  MAB & CR   &&   MAB & CR   &  MAB & CR   &&  MAB & CR   & MAB & CR   \\ \midrule
          \multicolumn{2}{l}{Full sample:}\\
          \partialinput{1}{5}{Simul/sim_SJ/bias_1.tex}
           \multicolumn{2}{l}{DDML methods:}\\
           \multicolumn{2}{l}{\it ~~Base learners}\\
          \partialinput{6}{19}{Simul/sim_SJ/bias_1.tex}
          \multicolumn{2}{l}{\it ~~Meta learners}\\
          \partialinput{20}{25}{Simul/sim_SJ/bias_1.tex}
    \midrule
    \bottomrule
    \end{tabular}%
  \begin{tablenotes}[para,flushleft]
  \scriptsize
  \item \textit{Notes:} The table reports median absolute bias (MAB) and coverage rate of a  95\% confidence interval (CR). We employ standard errors robust to heteroskedasticity.
  For comparison, we report the following full sample estimators: infeasible Oracle, OLS, PDS-Lasso with base and two different expanded sets of covariates. 
  DDML estimators use 20 folds for cross-fitting if $n=100$, and 5 folds if $n=1000$. Meta-learning approaches rely on all listed base learners.
  Results are based on 1,000 replications. Results for DGPs 4 and 5 can be found in Table~B.1 in the Appendix.
  \end{tablenotes}
\end{threeparttable}
\end{sidewaystable}

\section{The program}\label{sec:program}
In this section, we provide an overview of the \ddml package. We introduce the syntax and workflow for the main programs in Section~\ref{sec:syntax_ddml}. Section~\ref{sec:options} lists the options. Section~\ref{sec:syntax_qddml} covers the simplified one-line program \qddml. We provide an overview of supported machine learning programs in Section~\ref{sec:supported_programs}. Finally, Section~\ref{sec:replication} adds a note on how to ensure replication with \ddml.

\subsection{Syntax: \ddml}\label{sec:syntax_ddml}
The \texttt{ddml} estimation proceeds in four steps. 

\subsubsection*{Step 1: Initialize \ddml and select model.}

\begin{stsyntax}
ddml init {\it model}
    \optional{,
    mname({\it name})
    vars({\it varlist})
    kfolds({\it integer})
    fcluster({\it varname})
    foldvar({\it varlist})
    reps({\it integer})
    tabfold
    vars({\it varlist})
    }
\end{stsyntax}
\noindent where {\it model} selects between the Partially Linear Model (\texttt{partial}), the Interactive Model (\texttt{interactive}), the Partially Linear IV Model (\texttt{iv}), the Flexible Partially Linear IV Model (\texttt{fiv}), and the Interactive IV Model (\texttt{interactiveiv}). This step creates a persistent Mata object with the name provided by \stopt{mname}{name} in which model specifications and estimation results will be stored. The default name is {\it m0}.

At this stage, the user-specified folds for cross-fitting can be set via integer-valued Stata variables (see {\tt foldvar({\it varlist})}). By default, observations are randomly assigned to folds and {\tt kfolds({\it integer})} determines the number of folds (the default is 5). Cluster-randomized fold splitting is supported (see {\tt fcluster({\it varname})}). The user can also select the number of times to fully repeat the cross-fitting procedure (see {\tt rep({\it integer})}). 

\subsubsection*{Step 2: Add supervised machine learners for estimating conditional expectations.}

In this second step, we select the machine learning programs for estimating CEFs. 

\begin{stsyntax}
ddml {\it cond\_exp} 
    \optional{,
    mname({\it name})
    vname({\it varname})
    learner({\it name})
    vtype({\it string})
    predopt({\it string})
    }
    :
    {\it command} {\it depvar} {\it vars} 
    \optional{, 
    {\it cmdopt}
    }
\end{stsyntax}

\noindent where {\it cond\_exp} selects the conditional expectation to be estimated by the machine learning program \textit{command}. At least one learner is required for each conditional expectation. Table~\ref{tab:cond_exp} provides an overview of which conditional expectations are required by each model. The program \textit{command} is a supervised machine learning program such as \texttt{cvlasso} or \pystacked (see compatible programs in Section~\ref{sec:supported_programs}). The options {\it cmdopt} are specific to that program.

\begin{table}[H]
\newcommand{\xmark}{} 
    \centering
    \begin{tabular}{l|ccccc}\hline\hline
       {\it cond\_exp}      & \texttt{partial} &   \texttt{interactive} & \texttt{iv} & \texttt{fiv} &  \texttt{\revised{interactiveiv}} \\ \hline
       \verb+E[Y|X]+ &  \checkmark &  \xmark & \checkmark & \checkmark & \xmark\\
       \verb+E[Y|X,D]+ &\xmark & \checkmark &\xmark & \xmark & \xmark \\
       \verb+E[Y|X,Z]+ & \xmark & \xmark & \xmark & \xmark & \checkmark\\[.1cm]
       \verb+E[D|X]+ & \checkmark & \checkmark & \checkmark & \checkmark & \xmark \\
       \verb+E[D|Z,X]+ & \xmark & \xmark & \xmark & \checkmark & \checkmark  \\[.1cm]
       \verb+E[Z|X]+  & \xmark & \xmark & \checkmark & \xmark & \checkmark \\\hline\hline
    \end{tabular}
    \caption{The table lists the conditional expectations which need to be specified for each model.}
    \label{tab:cond_exp}
\end{table}

\subsubsection*{Step 3: Perform cross-fitting.}

This step implements the cross-fitting algorithm. Each learner is fit iteratively on training folds and out-of-sample predicted values are obtained. Cross-fitting is the most time-consuming step, as it involves fitting the selected machine learners repeatedly. 

\begin{stsyntax}
ddml crossfit 
    \optional{,
    mname({\it name})
    shortstack
    \revised{poolstack}
    \revised{nostdstack}
    \revised{finalest({\it name})}
    }
\end{stsyntax}

\subsubsection*{Step 4: Estimate causal effects.}

In the last step, we estimate the parameter of interest for all combination of learners added in Step 2.

\begin{stsyntax}
ddml estimate 
    \optional{,
    mname({\it name})
    robust
    cluster({\it varname})
    vce({\it vcetype})
    \revised{atet}
    trim
    \revised{shortstack}
    \revised{poolstack}
    \revised{finalest({\it name})}
    }
\end{stsyntax}

\noindent To report and post selected results, we can use {\tt ddml estimate} with the {\tt replay} option:

\begin{stsyntax}
ddml estimate 
    \optional{,
    replay
    mname({\it name})
    spec({\it integer or string})
    rep({\it integer or string})
    fulltable
    notable
    allest
    }
\end{stsyntax}

\subsubsection*{\revised{Utilities}}

\revised{{\tt ddml describe} provides information about the model setup and or results:}

\begin{stsyntax} \revisedcolor
ddml describe 
    \optional{,
    mname({\it name})
    sample
    learners
    crossfit
    estimates
    all
    }
\end{stsyntax}

\noindent \revised{{\tt ddml} stores many internal results on associate arrays, notably the various stacking weights.
These can be retrieved using {\tt ddml extract:}}

\begin{stsyntax}\revisedcolor
ddml extract
    \optional{{\it objectname} ,
    mname({\it name})
    show({\it name})
    ename({\it name})
    vname({\it varname})
    stata
    keys
    key1({\it string})
    key2({\it string})
    key3({\it string})
    subkey1({\it string})
    subkey2({\it string})
    }
\end{stsyntax}

\noindent \revised{{\tt ddml export} saves the estimated conditional expectations and other variables to a CSV file:}

\begin{stsyntax}\revisedcolor
ddml export {\it filename}
    \optional{,
    mname({\it name})
    addvars({\it varlist})
    }
\end{stsyntax}

\subsection{Options}\label{sec:options} 

\paragraph{Step 1 options: Initialization.}
\begin{description}
    \item[\stopt{mname}{{\it name}}] name of the DDML model. Allows to run multiple DDML models simultaneously. Defaults to {\it m0}.
    \item[\stopt{kfolds}{{\it integer}}] number of cross-fitting folds. The default is 5.
    \item[\stopt{fcluster}{{\it varname}}] cluster identifiers for cluster randomization of folds.
    \item[\stopt{foldvar}{{\it varlist}}] integer variables to specify custom folds (one per cross-fitting repetition).
    \item[\stopt{reps}{{\it integer}}] number of cross-fitting repetitions, i.e., how often the cross-fitting procedure is repeated on randomly generated folds.
    \item[{{\tt tabfold}}]  prints a table with frequency of observations by fold.
\end{description}

\paragraph{Step 2 options: Adding learners.}
\begin{description}
    \item[\stopt{vname}{{\it varname}}] name of the dependent variable in the reduced form estimation.  This is usually inferred from the {\it command} line but is mandatory for the {\tt fiv} model.
    \item[\stopt{learner}{{\it varname}}] optional name of the variable to be created.
    \item[\stopt{vtype}{{\it string}}] optional variable type of the variable to be created. Defaults to {\it double}.  {\it none} can be used to leave the type field blank. (Setting \stopt{vtype}{{\it none}} is required when using \ddml with {\tt rforest}.)
    \item[\stopt{predopt}{{\it varname}}]  {\tt predict} option to be used to get predicted values.  Typical values could be {\tt xb} or {\tt pr}. Default is blank.
\end{description}

\paragraph{Step 3 options: Cross-fitting.}
\begin{description}
    \item[{\ttfamily shortstack}]  asks for short-stacking to be used.  Short-stacking \revised{uses} the cross-fitted predicted values to obtain a weighted average of several base learners.
    \item[\revised{\ttfamily poolstack}]  \revised{asks for pooled stacking to be used. This is available only if {\tt pystacked} has been used for standard stacking in all equations.}
    \item[\revised{\ttfamily nostdstack}]  \revised{is used in conjunction with {\tt pystacked} and short-stacking; it tells {\tt pystacked} to generate the base learner predictions without the computationally-expensive additional step of obtaining the stacking weights.}
    \item[\revised{\stopt{finalest}{{\it name}}}] \revised{sets the final estimator for all stacking methods; the default {\it nnls1} is least squares without a constant and with the constraints that weights are non-negative and sum to one. Alternative final estimators include {\it singlebest} (use the minimum MSE base learner), {\it ols} (ordinary least squares) and {\it avg} (unweighted average of all base learners).}
\end{description}

\paragraph{Step 4 options: Estimation.}
\begin{description}
    \item[\stopt{spec}{{\it string}}] select specification. This can either be the specification number, {\it mse} for minimum-MSE specification (the default) or {\it ss} for short-stacking.
    \item[\stopt{rep}{{\it string}}] select cross-fitting repetitions. This can either be the cross-fit repetition number, {\it mn} for mean aggregation or {\it md} for median aggregation (the default). See Remark~2 for more information.
    \item[{\tt robust}] report SEs that are robust to the presence of arbitrary heteroskedasticity.
    \item[\stopt{cluster}{{\it varname}}] select cluster-robust variance-covariance estimator.
    \item[\stopt{vce}{{\it type}}] select variance-covariance estimator, e.g.\ \texttt{vce(hc3)} or \texttt{vce(cluster id)}. See \verb+help regress##vcetype+ for available options.
    \item[\stopt{trim}{{\it real}}] trimming of propensity scores for the Interactive and Interactive IV models. The default is 0.01 (that is, values below 0.01 and above 0.99 are set to 0.01 and 0.99, respectively).
    \item[{\tt atet}] report average treatment effect of the treated (default is ATE).
    \item[{\tt noconstant}] suppress constant term in the estimation stage (only relevant for partially linear models).
    \item[\revised{\ttfamily shortstack}, \revised{\ttfamily poolstack}, \revised{\stopt{finalest}{{\it name}}}] \revised{see above under {\tt ddml crossfit}.}
\end{description}
\revised{Re-fitting the final learner using {\tt ddml estimate} with stacking options is generally very fast because it does not require re-cross-fitting.}

\subsection{Short syntax: \qddml}\label{sec:syntax_qddml}
The \ddml package includes the wrapper program \qddml which provides a one-line syntax for estimating a \ddml model. The one-line syntax follows the syntax of \texttt{pdslasso} and \texttt{ivlasso} \citep{Ahrens2018}. The main restriction of \qddml compared to the more flexible multi-line syntax is that \qddml only allows for one user-specified machine learner (in addition to \texttt{regress}, which is added by default).

\revised{\qddml has integrated support for \pystacked, and \pystacked, is the default learner in all equations.
The syntax for \qddml options differs depending on whether \pystacked is used as the learner in each equation.}

\subsubsection*{\revised{Syntax when used with \pystacked}}
\begin{stsyntax} \revisedcolor
qddml
    \depvar\
    {\it treatment\_vars}
    \textnormal{(}{\it controls}\textnormal{)}
    \textnormal{(}{\it treatment\_vars}=excluded\_instruments\textnormal{)}
    \optional{
    pystacked(string)
    pystacked\_y(string)
    pystacked\_d(string)
    pystacked\_z(string)
    shortstack
    stdstack
    poolstack
    finalest(name)
    mname(name)
    {\it options}
     }
\end{stsyntax}
\noindent \revised{The \pystacked option sets the options for all the conditional expectations estimated by \pystacked; the \texttt{\_y}, \texttt{\_d} and \texttt{\_z} variants control the options sent to the corresponding conditional expectation estimations.
Other options are as in \ddml.}

\subsubsection*{\revised{Syntax when used with other learners}}
\begin{stsyntax} \revisedcolor
qddml
    \depvar\
    {\it treatment\_vars}
    \textnormal{(}{\it controls}\textnormal{)}
    \textnormal{(}{\it treatment\_vars}=excluded\_instruments\textnormal{)}
    \optional{
    cmd(string)
    ycmd(string)
    dcmd(string)
    zcmd(string)
    *cmdopt(string)
    *vtype(name)
    *predopt(name)
    {\it options}
     }
\end{stsyntax}
\noindent \revised{The \texttt{cmd} option sets the options for all the conditional expectations estimated by \pystacked; the \texttt{ycmd}, \texttt{dcmd} and \texttt{zcmd} variants control the options sent to the corresponding conditional expectation estimations.} 
\revised{The \texttt{cmdopt} option can be used either to set the options for all equations, or, by replacing the asterisk with y, d or z, set the options for the corresponding conditional expectation estimation.
Other options are as in \ddml.}

%
%

\subsection{Supported machine learning programs}\label{sec:supported_programs}
\ddml is compatible with any supervised ML program in Stata that supports the typical ``\texttt{reg\,\,y\,\,x}'' syntax, comes with a post-estimation \stcmd{predict} and supports \stcmd{if} statements. We have tested \ddml with the following programs:
\begin{itemize}[nosep]
    \item \revised{\pystacked facilitates the stacking of a wide range of machine learners including regularized regression, random forests, support vector machines, gradient boosted trees and feed-forward neural nets using Python's \emph{scikit-learn} \citep{Ahrens2022,scikit-learn,sklearn_api}. In addition, \pystacked can also be used as a front-end to fit individual machine learners. \ddml has integrated support for \pystacked and is the recommended default learner.}
    \item \texttt{lassopack} implements regularized regression, e.g.\ lasso, ridge, elastic net \citep{Ahrens2019}.
    \item \texttt{rforest} is a random forest wrapper for WEKA \citep{Schonlau2020,frank2009weka}.
    \item \texttt{svmachines} allows for the estimation of support vector machines using \texttt{libsvm} \citep{chang2011libsvm,Guenther2018}.
    \item The program \texttt{parsnip} of the package \texttt{mlrtime} provides access to R's \emph{parsnip} machine learning library through \texttt{rcall} \citep{Huntington2021,haghish2019seamless}. Using \texttt{parsnip} requires the installation of the supplementary wrapper program \texttt{parsnip2}.\footnote{Available from {\tt\url{https://github.com/aahrens1/parsnip2}.}}
\end{itemize}

\noindent Stata programs that are currently not supported can be added relatively easily using wrapper programs (see \texttt{parsnip2} for an example).

\subsection{Inspecting results and replication}\label{sec:replication}
In this section we discuss how to ensure replicability when using \ddml. We also discuss some tools available for tracing replication failures.
First, however, we briefly describe how \ddml stores results.

\ddml stores estimated conditional expectations in Stata's memory as Stata variables. These variables can be inspected, graphed and summarized in the usual way.
Fold ID variables are also stored as Stata variables (by default named \verb+m0_fid_+\textit{r}, where \verb+m0+ is the default model name and \textit{r} is the cross-fitting repetition). \ddml models are stored on Mata \texttt{struct}s and using Mata's associative arrays. Specifically, the \ddml model created by \ddml \texttt{init} is an \texttt{mStruct}, and information relating to the estimation of conditional expectations are stored in \texttt{eStruct}s. Results relating to the overall model estimation are stored in associative arrays that live in the \texttt{mStruct}, and results relating to the estimation of conditional expectations are stored in associative arrays that live in the corresponding \texttt{eStruct}s.

Replication tips:
\begin{itemize} 
    \item Set the Stata seed before {\tt ddml init}. This \revised{ensures} that the same random fold variable is used for a given data set. 
    \item Using the same fold variable alone is usually not sufficient to ensure replication, since many machine learning algorithms involve randomization. That said, note that the fold variable is stored in memory and can be reused for subsequent estimations via the \stopt{foldvar}{varlist} option.
    \item Replication of \ddml results may require additional steps with some programs that rely on randomization in other software environments, e.g., R or Python. \pystacked uses a Python seed generated in Stata. Thus, when \ddml is used with \pystacked, setting the seed before {\tt ddml init} also guarantees that the same Python seed underlies the stacking estimation. Other programs relying on randomization outside of Stata might not behave in the same way. Thus, when using other programs, check the help files for options to set external random seeds. Try estimating each of the individual learners on the entire sample to see what settings need to be passed to them for their results to replicate. 
    \item Beware of changing samples. Fold splits or learner idiosyncracies may mean that sample sizes vary slightly across learners, estimation samples and/or cross-fitting repetitions.
    \item The \ddml \texttt{export} utility can be used to export the estimated conditional expectations, fold variables and sample indicators to a CSV format file for examination and comparison in other software environments.
\end{itemize}

\section{Applications}\label{sec:application}
We demonstrate the \texttt{ddml} workflow using two applications. In Section~\ref{sec:401k}, we apply the DDML estimator to estimate the effect of 401(k) eligibility on financial wealth following \citet{poterba1995}. We focus on the Partially Linear Model for the sake of brevity, but provide code that demonstrates the use of \ddml with the Interactive Model, Partially Linear IV Model and Interactive IV Model using the same application in Appendix~C. Additional examples can also be found in the help file. 
Based on \citet{berry1995automobile}, we show in Section~\ref{sec:ivhd_application} how to employ \ddml for the estimation of the Flexible Partially Linear IV Model which allows  both for flexibly controlling for confounding factors using high-dimensional function approximation of confounding factors and for estimation of optimal instrumental variables. 

\subsection{401(k) and financial wealth}\label{sec:401k}
The data consists of $n=9915$ households from the 1991 SIPP. The application is originally due to \citet{poterba1995}, but has been revisited by \citet{Belloni2017}, \citet{Chernozhukov2018}, and \citet{wuthrich2021}, among others. Following previous studies, we include the control variables age, income, years of education, family size, as well as indicators for martial status, two-earner status, benefit pension status, IRA participation, and home ownership. The outcome is net financial assets and the treatment is eligibility to enroll for the 401(k) pension plan. 

We load the data and define three globals for outcome, treatment and control variables. We then proceed in the four steps outlined in Section~\ref{sec:syntax_ddml}. 

\begin{stlog}
. use "sipp1991.dta", clear
{\smallskip}
. global Y net_tfa
{\smallskip}
. global X age inc educ fsize marr twoearn db pira hown
{\smallskip}
. global D e401
{\smallskip}

\end{stlog}

\subsubsection*{Step 1: Initialize \texttt{ddml} model.} We initialize the \ddml model and select the Partially Linear Model in \eqref{eq:plm}. Before initialization, we set the seed to ensure replication. This should always be done before \texttt{ddml init}, which executes the random fold assignment. In this example, we opt for four folds to ensure the readability of some of the output shown below, although we recommend considering a larger number of folds in practice. 

\begin{stlog}
. set seed 123
{\smallskip}
. ddml init partial, kfolds(4)

\end{stlog}

\subsubsection*{Step 2: Add supervised machine learners for estimating conditional expectations.} In this step, we specify which machine learning programs should be used for the estimation of the conditional expectations $E[Y\vert \bm{X}]$ and $E[D\vert \bm{X}]$. For each conditional expectation, at least one learner is required. \revised{For illustrative purposes, we consider \texttt{regress} for linear regression, \texttt{pystacked} with the \texttt{m(lassocv)} option for cross-validated lasso (as an example of how to use \texttt{pystacked} to estimate a single learner) and \texttt{rforest} for random forests.} When using \texttt{rforest}, we need to add the option \texttt{vtype(none)} since the post-estimation \texttt{predict} command of \texttt{rforest} does not support variable types. 

\begin{stlog}
. *** add learners for E[Y|X]
. ddml E[Y|X]: reg $Y $X
Learner Y1_reg added successfully.
{\smallskip}
. ddml E[Y|X]: pystacked $Y c.($X)\#\#c.($X), type(reg) m(lassocv)
Learner Y2_pystacked added successfully.
{\smallskip}
. ddml E[Y|X], vtype(none): rforest $Y $X, type(reg)
Learner Y3_rforest added successfully.
{\smallskip}
. *** add learners for E[D|X]
. ddml E[D|X]: reg $D $X
Learner D1_reg added successfully.
{\smallskip}
. ddml E[D|X]: pystacked $D c.($X)\#\#c.($X), type(reg) m(lassocv)
Learner D2_pystacked added successfully.
{\smallskip}
. ddml E[D|X], vtype(none): rforest $D $X, type(reg)
Learner D3_rforest added successfully.

\end{stlog}

The flexible \ddml syntax allows \revised{specification of different} sets of covariates for different learners. This flexibility can be useful as, for example, linear learners such as the lasso might perform better if interactions are provided as inputs, whereas tree-based methods such as random forests may detect certain interactions in a data-driven way. Here, we use interactions and second-order polynomials for \revised{the cross-validated lasso,} but not for the other learners.

This application has only one treatment variable, but \ddml does support multiple treatment variables. To add a second treatment variable, we would simply add a statement such as \texttt{ddml E[D|X]: reg D2 \$X} where \texttt{D2} would be the name of the second treatment variable. An example with two treatments is provided in the help file.

The auxiliary \texttt{ddml} sub-command \texttt{describe} allows \revised{us} to verify that the learners were correctly registered:

\begin{stlog}
. ddml describe
{\smallskip}
Model:                  partial, crossfit folds k=4, resamples r=1
Mata global (mname):    m0
Dependent variable (Y): net_tfa
 net_tfa learners:      Y1_reg Y2_pystacked Y3_rforest
D equations (1):        e401
 e401 learners:         D1_reg D2_pystacked D3_rforest
{\smallskip}

\end{stlog}

\subsubsection*{Step 3: Perform cross-fitting.} The third step performs cross-fitting, which is the most time-intensive process. The \texttt{shortstack} option enables the short-stacking algorithm of Section~\ref{eq:stacking}.

\begin{stlog}
. ddml crossfit, shortstack
Cross-fitting E[y|X] equation: net_tfa
Cross-fitting fold 1 2 3 4 ...completed cross-fitting...completed short-stacking
Cross-fitting E[D|X] equation: e401
Cross-fitting fold 1 2 3 4 ...completed cross-fitting...completed short-stacking

\end{stlog}

\revisedcolor

Six variables are created and stored in memory which correspond to the six learners specified in the previous step. These variables are called \verb+Y1_reg_1+, \verb+Y2_pystacked_1_1+, \verb+Y3_rforest_1+, \verb+D1_reg_1+, \verb+D2_pystacked_1_1+ and \verb+D3_rforest_1+. \texttt{Y} and \texttt{D} indicate outcome and treatment variable. The index \texttt{1} to \texttt{3} is a learner counter. \texttt{reg}, \texttt{pystacked} and \texttt{rforest} correspond to the names of the commands used. The \verb+_1+ suffix  indicates the cross-fitting repetition. The additional \verb+_1+ in the case of \verb+D2_pystacked_1_1+ indicates the learner number (there is only a single \texttt{pystacked} learner).

\revisedcoloroff

After cross-fitting, we can inspect the mean-squared prediction errors by fold and learner:
\begin{stlog}
. ddml desc, crossfit
{\smallskip}
Model:                  partial, crossfit folds k=4, resamples r=1
Mata global (mname):    m0
Dependent variable (Y): net_tfa
 net_tfa learners:      Y1_reg Y2_pystacked Y3_rforest
D equations (1):        e401
 e401 learners:         D1_reg D2_pystacked D3_rforest
{\smallskip}
Crossfit results (detail):
                                     All    By fold:
Cond. exp.  Learner      rep         MSE        1         2         3         4 
net_tfa    shortstack     1       1.5e+09   1.8e+09   1.4e+09   1.4e+09   1.5e+09
e401       shortstack     1          0.18      0.17      0.17      0.18      0.18
{\smallskip}

\end{stlog}

\subsubsection*{Step 4: Estimate causal effects.}

In this final step, we obtain the causal effect estimates. Since we requested short-stacking in Step~3, \texttt{ddml} shows \revised{the} short-stacking result which relies on the cross-fitted values of each base learner. In addition, the specification that corresponds to the minimum-MSE learners is listed at the beginning of the output.

\begin{stlog}
. ddml estimate, robust  
{\smallskip}
{\smallskip}
Model:                  partial, crossfit folds k=4, resamples r=1
Mata global (mname):    m0
Dependent variable (Y): net_tfa
 net_tfa learners:      Y1_reg Y2_pystacked Y3_rforest
D equations (1):        e401
 e401 learners:         D1_reg D2_pystacked D3_rforest
{\smallskip}
DDML estimation results:
spec  r     Y learner     D learner         b        SE 
 mse  1  Y2_pystacked  D2_pystacked  9788.291 (1339.797)
  ss  1  [shortstack]          [ss]  9748.788 (1331.364)
mse = minimum MSE specification for that resample.
{\smallskip}
Shortstack DDML model
y-E[y|X]  = y-Y_net_tfa_ss_1                       Number of obs   =      9915
D-E[D|X]  = D-D_e401_ss_1 
\HLI{13}{\TOPT}\HLI{64}
             {\VBAR}               Robust
     net_tfa {\VBAR} Coefficient  std. err.      z    P>|z|     [95\% conf. interval]
\HLI{13}{\PLUS}\HLI{64}
        e401 {\VBAR}   9748.788   1331.364     7.32   0.000     7139.362    12358.21
       _cons {\VBAR}   92.48199   534.8262     0.17   0.863    -955.7581    1140.722
\HLI{13}{\BOTT}\HLI{64}
Stacking final estimator: nnls1
{\smallskip}
{\smallskip}

\end{stlog}

\revisedcolor

Since we have specified three learners per conditional expectation, there are in total 9 specifications relying on the base learners (since we can combine  \verb+Y1_reg_1+, \verb+Y2_pystacked_1+ and \verb+Y3_rforest_1+ with \verb+D1_reg_1+, \verb+D2_pystacked_1+ and \verb+D3_rforest_1+). To get all results, we add the \texttt{allcombos} option:

\revisedcoloroff

\begin{stlog}
. ddml estimate, robust allcombos
{\smallskip}
{\smallskip}
Model:                  partial, crossfit folds k=4, resamples r=1
Mata global (mname):    m0
Dependent variable (Y): net_tfa
 net_tfa learners:      Y1_reg Y2_pystacked Y3_rforest
D equations (1):        e401
 e401 learners:         D1_reg D2_pystacked D3_rforest
{\smallskip}
DDML estimation results:
spec  r     Y learner     D learner         b        SE 
   1  1        Y1_reg        D1_reg  5986.657 (1523.694)
   2  1        Y1_reg  D2_pystacked  9563.875 (1389.172)
   3  1        Y1_reg    D3_rforest  8659.141 (1258.982)
   4  1  Y2_pystacked        D1_reg  9175.519 (1371.065)
*  5  1  Y2_pystacked  D2_pystacked  9788.291 (1339.797)
   6  1  Y2_pystacked    D3_rforest  8497.214 (1200.268)
   7  1    Y3_rforest        D1_reg  9044.071 (1485.073)
   8  1    Y3_rforest  D2_pystacked 10103.630 (1426.134)
   9  1    Y3_rforest    D3_rforest  9528.449 (1293.485)
  ss  1  [shortstack]          [ss]  9748.788 (1331.364)
* = minimum MSE specification for that resample.
{\smallskip}
Shortstack DDML model
y-E[y|X]  = y-Y_net_tfa_ss_1                       Number of obs   =      9915
D-E[D|X]  = D-D_e401_ss_1 
\HLI{13}{\TOPT}\HLI{64}
             {\VBAR}               Robust
     net_tfa {\VBAR} Coefficient  std. err.      z    P>|z|     [95\% conf. interval]
\HLI{13}{\PLUS}\HLI{64}
        e401 {\VBAR}   9748.788   1331.364     7.32   0.000     7139.362    12358.21
       _cons {\VBAR}   92.48199   534.8262     0.17   0.863    -955.7581    1140.722
\HLI{13}{\BOTT}\HLI{64}
Stacking final estimator: nnls1
{\smallskip}
{\smallskip}

\end{stlog}

We can use the \stopt{spec}{string} option to select among the listed specifications. {\it string} is either the specification number, \revised{\texttt{ss}, \texttt{st} or \texttt{ps},  to get the short-stacking, standard stacking or pooled stacking specification, respectively,} or \texttt{mse} for the specification corresponding to the minimal MSPE. In the example above, \texttt{spec(1)} reports in full the specification using \texttt{regress} for estimating both $E[Y\vert \bm{X}]$ and $E[D\vert \bm{X}]$. The \stopt{spec}{string} option can be provided either in combination with \texttt{allcombos}, or after estimation in combination with the {\tt replay} option, for example:

\begin{stlog}
. ddml estimate, spec(1) replay
{\smallskip}
{\smallskip}
Model:                  partial, crossfit folds k=4, resamples r=1
Mata global (mname):    m0
Dependent variable (Y): net_tfa
 net_tfa learners:      Y1_reg Y2_pystacked Y3_rforest
D equations (1):        e401
 e401 learners:         D1_reg D2_pystacked D3_rforest
{\smallskip}
DDML estimation results:
spec  r     Y learner     D learner         b        SE 
   1  1  Y2_pystacked  D2_pystacked  5986.657 (1523.694)
  ss  1  [shortstack]          [ss]  9748.788 (1331.364)
mse = minimum MSE specification for that resample.
{\smallskip}
DDML model, specification 1
y-E[y|X]  = y-Y1_reg_1                             Number of obs   =      9915
D-E[D|X]  = D-D1_reg_1 
\HLI{13}{\TOPT}\HLI{64}
             {\VBAR}               Robust
     net_tfa {\VBAR} Coefficient  std. err.      z    P>|z|     [95\% conf. interval]
\HLI{13}{\PLUS}\HLI{64}
        e401 {\VBAR}   5986.657   1523.694     3.93   0.000     3000.271    8973.042
       _cons {\VBAR}   10.74705   561.2911     0.02   0.985    -1089.363    1110.857
\HLI{13}{\BOTT}\HLI{64}
{\smallskip}
{\smallskip}

\end{stlog}



\subsubsection*{Manual final estimation.}
In the background, \texttt{ddml estimate} regresses \verb+Y1_reg_1+ against \verb+D1_reg_1+ with a constant. We can verify this manually:

\begin{stlog}
. gen double Y1_resid = $Y - Y1_reg
{\smallskip}
. gen double D1_resid = $D - D1_reg
{\smallskip}
. reg Y1_resid D1_resid,  robust
{\smallskip}
Linear regression                               Number of obs     =      9,915
                                                F(1, 9913)        =      15.44
                                                Prob > F          =     0.0001
                                                R-squared         =     0.0023
                                                Root MSE          =      55891
{\smallskip}
\HLI{13}{\TOPT}\HLI{64}
             {\VBAR}               Robust
    Y1_resid {\VBAR} Coefficient  std. err.      t    P>|t|     [95\% conf. interval]
\HLI{13}{\PLUS}\HLI{64}
    D1_resid {\VBAR}   5986.657   1523.694     3.93   0.000     2999.906    8973.407
       _cons {\VBAR}   10.74705   561.2911     0.02   0.985    -1089.498    1110.992
\HLI{13}{\BOTT}\HLI{64}

\end{stlog}

Manual estimation using {\ttfamily regress} allows the use of {\ttfamily regress}'s postestimation tools.

%
%

%
%
%

\subsubsection*{Stacking}

We next demonstrate DDML with stacking. To this end, we exploit the stacking regressor implemented in \pystacked. \pystacked allows \revised{combining} multiple base learners with learner-specific settings and covariates into a final meta learner. The learners are separated by \verb+||+. \stopt{method}{name} selects the learner, \stopt{xvars}{varlist} specifies learner-specific covariates (overwritting the default covariates \verb+$X+) and \stopt{opt}{string} passes options to the learners. In this example, we use OLS, cross-validated lasso and ridge, random forests and gradient boosting. We furthermore use parallelization with 5 cores. A detailed explanation of the \pystacked syntax can be found in \citet{Ahrens2022}.

\begin{stlog}
. *** add learners for E[Y|X]
. ddml E[Y|X]: pystacked $Y $X                                             || ///
>     method(ols)                                                          || ///
>     m(lassocv) xvars(c.($X)\#\#c.($X))                                     || ///
>     m(ridgecv) xvars(c.($X)\#\#c.($X))                                     || ///
>     m(rf) pipe(sparse) opt(max_features(5))                              || ///
>     m(gradboost) pipe(sparse) opt(n_estimators(250) learning_rate(0.01)) ,  ///
>     njobs(5)
Learner Y1_pystacked added successfully.
{\smallskip}
. *** add learners for E[D|X]
. ddml E[D|X]: pystacked $D $X                                             || ///
>     method(ols)                                                          || ///
>     m(lassocv) xvars(c.($X)\#\#c.($X))                                     || ///
>     m(ridgecv) xvars(c.($X)\#\#c.($X))                                     || ///
>     m(rf) pipe(sparse) opt(max_features(5))                              || ///
>     m(gradboost) pipe(sparse) opt(n_estimators(250) learning_rate(0.01)) ,  ///
>     njobs(5)
Learner D1_pystacked added successfully.

\end{stlog}

\noindent After cross-fitting, we \revised{retrieve the cross-fitted MSPE using \texttt{ddml extract} with \texttt{show(mse)} or examine the stacking weights using \texttt{stweights}}:
\begin{stlog}
. qui ddml crossfit 
{\smallskip}
. ddml extract, show(stweights)
{\smallskip}
mean stacking weights across folds/resamples for D1_pystacked (e401)
final stacking estimator: nnls1
               learner  mean_weight        rep_1
      ols            1    .01557419    .01557419
  lassocv            2    .10077907    .10077907
  ridgecv            3    .43674242    .43674242
       rf            4    .02946916    .02946916
gradboost            5    .41743516    .41743516
{\smallskip}
mean stacking weights across folds/resamples for Y1_pystacked (net_tfa)
final stacking estimator: nnls1
               learner  mean_weight        rep_1
      ols            1    .09662631    .09662631
  lassocv            2    .46475744    .46475744
  ridgecv            3    .32388159    .32388159
       rf            4    .09392877    .09392877
gradboost            5     .0145518     .0145518
{\smallskip}

\end{stlog}

Finally, \revised{in the estimation stage,} we retrieve the results of DDML with stacking:

\begin{stlog}
. ddml estimate, robust
{\smallskip}
{\smallskip}
Model:                  partial, crossfit folds k=4, resamples r=1
Mata global (mname):    m0
Dependent variable (Y): net_tfa
 net_tfa learners:      Y1_pystacked
D equations (1):        e401
 e401 learners:         D1_pystacked
{\smallskip}
DDML estimation results:
spec  r     Y learner     D learner         b        SE 
  st  1  Y1_pystacked  D1_pystacked  9406.385 (1300.170)
{\smallskip}
Stacking DDML model
y-E[y|X]  = y-Y1_pystacked_1                       Number of obs   =      9915
D-E[D|X]  = D-D1_pystacked_1 
\HLI{13}{\TOPT}\HLI{64}
             {\VBAR}               Robust
     net_tfa {\VBAR} Coefficient  std. err.      z    P>|z|     [95\% conf. interval]
\HLI{13}{\PLUS}\HLI{64}
        e401 {\VBAR}   9406.385    1300.17     7.23   0.000     6858.099    11954.67
       _cons {\VBAR}   199.9921   535.7477     0.37   0.709    -850.0541    1250.038
\HLI{13}{\BOTT}\HLI{64}
Stacking final estimator: nnls1
{\smallskip}
{\smallskip}

\end{stlog}


\revisedcolor

The DDML-specific stacking approaches of short-stacking and pooled stacking can be requested at either the cross-fitting or estimation steps. Re-fitting the final learner at the estimation step allows us to avoid repeating the computationally-intensive cross-fitting. Here, we request short-stacking and pooled stacking but using the single-best base learner.

\begin{stlog}
. ddml estimate, robust shortstack poolstack finalest(singlebest)
{\smallskip}
{\smallskip}
Model:                  partial, crossfit folds k=4, resamples r=1
Mata global (mname):    m0
Dependent variable (Y): net_tfa
 net_tfa learners:      Y1_pystacked
D equations (1):        e401
 e401 learners:         D1_pystacked
{\smallskip}
DDML estimation results:
spec  r     Y learner     D learner         b        SE 
  st  1  Y1_pystacked  D1_pystacked  9406.385 (1300.170)
  ss  1  [shortstack]          [ss]  9847.256 (1339.419)
  ps  1   [poolstack]          [ps]  9788.291 (1339.797)
{\smallskip}
Shortstack DDML model
y-E[y|X]  = y-Y_net_tfa_ss_1                       Number of obs   =      9915
D-E[D|X]  = D-D_e401_ss_1 
\HLI{13}{\TOPT}\HLI{64}
             {\VBAR}               Robust
     net_tfa {\VBAR} Coefficient  std. err.      z    P>|z|     [95\% conf. interval]
\HLI{13}{\PLUS}\HLI{64}
        e401 {\VBAR}   9847.256   1339.419     7.35   0.000     7222.044    12472.47
       _cons {\VBAR}   78.69447   534.6452     0.15   0.883    -969.1909     1126.58
\HLI{13}{\BOTT}\HLI{64}
Stacking final estimator: singlebest
{\smallskip}
{\smallskip}

\end{stlog}

Short-stacking is computationally substantially faster than regular stacking (or pooled stacking) because it avoids the cross-validation within cross-fit folds. Below, we disable regular stacking with the \texttt{nostdstack} option in the cross-fitting stage. In this example, where we use parallelization with 5 cores, the run time is only 50.7 seconds DDML with short-stakcing compared to 93.0 seconds for DDML with regular stacking.

\begin{stlog}
. qui ddml crossfit, shortstack nostdstack
{\smallskip}
. ddml estimate, robust
{\smallskip}
{\smallskip}
Model:                  partial, crossfit folds k=4, resamples r=1
Mata global (mname):    m0
Dependent variable (Y): net_tfa
 net_tfa learners:      Y1_pystacked
D equations (1):        e401
 e401 learners:         D1_pystacked
{\smallskip}
DDML estimation results:
spec  r     Y learner     D learner         b        SE 
  ss  1  [shortstack]          [ss]  9602.257 (1300.825)
{\smallskip}
Shortstack DDML model
y-E[y|X]  = y-Y_net_tfa_ss_1                       Number of obs   =      9915
D-E[D|X]  = D-D_e401_ss_1 
\HLI{13}{\TOPT}\HLI{64}
             {\VBAR}               Robust
     net_tfa {\VBAR} Coefficient  std. err.      z    P>|z|     [95\% conf. interval]
\HLI{13}{\PLUS}\HLI{64}
        e401 {\VBAR}   9602.257   1300.825     7.38   0.000     7052.686    12151.83
       _cons {\VBAR}   83.96648   533.9871     0.16   0.875    -962.6289    1130.562
\HLI{13}{\BOTT}\HLI{64}
Stacking final estimator: 
{\smallskip}
{\smallskip}

\end{stlog}

\subsubsection*{One-line syntax.}
\qddml provides a simplified and convenient one-line syntax. The main constraint of \qddml is that it only allows for one user-specified learner. The default learner is \texttt{pystacked}, which by default uses OLS, cross-validated lasso and gradient boosting as default learners. The \texttt{pystacked} base learners as well as non-\texttt{pystacked} commands can be  be modified via various options. Below we use \texttt{qddml} with \texttt{pystacked}'s default base learners. We omit the output for the sake of brevity.

\begin{stlog}
. qui qddml $Y $D (c.($X)\#\#c.($X)), model(partial) kfolds(4) robust  
{\smallskip}

\end{stlog}

\revisedcoloroff

\subsection{The market for automobiles}\label{sec:ivhd_application}
For this demonstration, we follow \citet{chernozhukov2015} who estimate a stylized demand model using instrumental variables based on the data from \citet{berry1995automobile}. The authors of the original study estimate the effect of prices on the market share of automobile models in a given year ($n=2217$). The controls are product characteristics (a constant, air conditioning dummy, horsepower divided by weight, miles per dollar, vehicle size). To account for endogenous prices, \citet{berry1995automobile} suggest exploiting other products' characteristics as instruments. Following \citet{chernozhukov2015}, we define the baseline set of instruments as the sum over all other products' characteristics, calculated separately for own-firm and other-firm products, which yields 10 baseline instruments. \citet{chernozhukov2015} also construct an augmented set of instruments, including first-order interactions, squared and cubic terms. In the analysis below, we extend \citet{chernozhukov2015} by applying DDML with stacking and a diverse set of learners including OLS, lasso, ridge, random forest and gradient boosted trees. 
We use the augmented set of controls for all base learners and OLS, which we include for reference. 

We load and prepare the data:
\begin{stlog}
. use BLP_CHS.dta, clear
{\smallskip}
. global Y y
{\smallskip}
. global D price
{\smallskip}
. global Xbase hpwt air mpd space
{\smallskip}
. global Xaug augX*
{\smallskip}
. global Zbase Zbase*
{\smallskip}
. global Zaug Zaug*
{\smallskip}

\end{stlog} 

\paragraph{Step 1: Initialize \texttt{ddml} model.}

\begin{stlog}
. set seed 123
{\smallskip}
. ddml init fiv, kfolds(4) reps(5)

\end{stlog} 

Note that in the \texttt{ddml init} step, we include the option \texttt{reps(5)} which will result in running the full cross-fitting procedure five times, each with a different random split of the data. Replicating the procedure multiple times allows us to gauge the impact of randomness due to the random splitting of the data into subsamples.

\paragraph{Step 2: Add supervised machine learners for estimating conditional expectations.}

Estimation of a \texttt{fiv} model requires us to add learners for $E[Y|\bm{X}]$, $E[D|\bm{X},\bm{Z}]$ and $E[D|\bm{X}]$. Compared to the other models supported by \ddml, there is one complication that arises because, in order to estimate $E[D|\bm{X}]$, we exploit fitted values of $E[D|\bm{X},\bm{Z}]$ to impose LIE-compliance. Since these fitted values have not yet been generated, we use the placeholder \verb+{D}+ that in the cross-fitting stage will be internally replaced with estimates of $E[D|\bm{X},\bm{Z}]$. We use the \stopt{learner}{string} option to match one learner for $E[D|\bm{X}]$ with a learner for $E[D|\bm{X},\bm{Z}]$, and \stopt{vname}{varname} to indicate the name of the treatment variable. 

\begin{stlog}
. *** add learners for E[Y|X] 
. ddml E[Y|X], learner(Ypystacked): pystacked $Y $Xaug                    || ///
>     method(ols) xvars($Xbase)                                           || ///
>     m(lassocv)                                                          || ///
>     m(ridgecv)                                                          || ///
>     m(rf) opt(n_estimators(200) max_features(None))                     || ///
>     m(rf) opt(n_estimators(200) max_features(10))                       || ///
>     m(rf) opt(n_estimators(200) max_features(5))                        || ///
>     m(gradboost) opt(n_estimators(800) learning_rate(0.01))             || ///
>     m(gradboost) opt(n_estimators(800) learning_rate(0.1))              || ///
>     m(gradboost) opt(n_estimators(800) learning_rate(0.3))              ,  ///
>     njobs(4)
Learner Ypystacked added successfully.
{\smallskip}
. ddml E[D|X,Z], learner(Dpystacked): pystacked $D $Xaug $Zaug            || ///
>     method(ols) xvars($Xbase $Zbase)                                    || ///
>     m(lassocv)                                                          || ///
>     m(ridgecv)                                                          || ///
>     m(rf) opt(n_estimators(200) max_features(None))                     || ///
>     m(rf) opt(n_estimators(200) max_features(10))                       || ///
>     m(rf) opt(n_estimators(200) max_features(5))                        || ///
>     m(gradboost) opt(n_estimators(800) learning_rate(0.01))             || ///
>     m(gradboost) opt(n_estimators(800) learning_rate(0.1))              || ///
>     m(gradboost) opt(n_estimators(800) learning_rate(0.3))              ,  ///
>     njobs(4)
Learner Dpystacked added successfully.
{\smallskip}
. ddml E[D|X], mname(m0) learner(Dpystacked) vname($D):                   ///
>                                                 pystacked {\lbr}D{\rbr} $Xaug     || ///
>     method(ols) xvars($Xaug)                                            || ///
>     m(lassocv)                                                          || ///
>     m(ridgecv)                                                          || ///
>     m(rf) opt(n_estimators(200) max_features(None))                     || ///
>     m(rf) opt(n_estimators(200) max_features(10))                       || ///
>     m(rf) opt(n_estimators(200) max_features(5))                        || ///
>     m(gradboost) opt(n_estimators(800) learning_rate(0.01))             || ///
>     m(gradboost) opt(n_estimators(800) learning_rate(0.1))              || ///
>     m(gradboost) opt(n_estimators(800) learning_rate(0.3))              ,  ///
>     njobs(4)
Learner Dpystacked_h added successfully.

\end{stlog} 

\paragraph{Steps 3-4: Perform cross-fitting (output omitted) and estimate causal effects.}



\begin{stlog}
. qui ddml crossfit
{\smallskip}
. ddml estimate, robust
{\smallskip}
{\smallskip}
Model:                  fiv, crossfit folds k=4, resamples r=5
Mata global (mname):    m0
Dependent variable (Y): y
 y learners:            Ypystacked
D equations (1):        price
 price learners:        Dpystacked
{\smallskip}
DDML estimation results:
spec  r     Y learner     D learner         b        SE     DH learner
  st  1    Ypystacked    Dpystacked    -0.108    (0.010)  Dpystacked_h
  st  2    Ypystacked    Dpystacked    -0.123    (0.010)  Dpystacked_h
  st  3    Ypystacked    Dpystacked    -0.110    (0.011)  Dpystacked_h
  st  4    Ypystacked    Dpystacked    -0.127    (0.011)  Dpystacked_h
  st  5    Ypystacked    Dpystacked    -0.132    (0.012)  Dpystacked_h
{\smallskip}
Mean/med    Y learner     D learner         b        SE     DH learner
  st mn    Ypystacked    Dpystacked    -0.120    (0.014)  Dpystacked_h
  st md    Ypystacked    Dpystacked    -0.123    (0.015)  Dpystacked_h
{\smallskip}
Median over 5 stacking resamples
y-E[y|X]  = y-Ypystacked                           Number of obs   =      2217
E[D|X,Z]  = D-Dpystacked 
E[D{\caret}|X]   = Dpystacked_h
Orthogonalized D = D - E[D{\caret}|X]; optimal IV = E[D|X,Z] - E[D{\caret}|X].
\HLI{13}{\TOPT}\HLI{64}
             {\VBAR}               Robust
           y {\VBAR} Coefficient  std. err.      z    P>|z|     [95\% conf. interval]
\HLI{13}{\PLUS}\HLI{64}
       price {\VBAR}  -.1230112   .0152603    -8.06   0.000    -.1529209   -.0931016
\HLI{13}{\BOTT}\HLI{64}
Stacking final estimator: nnls1
{\smallskip}
Summary over 5 resamples:
       D eqn      mean       min       p25       p50       p75       max
       price     -0.1199   -0.1318   -0.1271   -0.1230   -0.1101   -0.1075
{\smallskip}

\end{stlog} 

\paragraph{Manual final estimation.}
We can obtain the final estimate manually. To this end, we construct the instrument as $\hat{E}[D|X,Z]-\hat{E}[D|X]$ and the residualized endogenous regressor as $D-\hat{E}[D|X]$. The residualized dependent variable is saved in memory. Here we obtain the estimate from the first cross-fitting replication. We could obtain the estimate for replication $r$ by changing the ``\texttt{\_1}'' to ``\texttt{\_}$r$''. 

\begin{stlog}
. gen double Y1_resid = $Y - Ypystacked_1
{\smallskip}
. gen double dtilde = $D - Dpystacked_h_1
{\smallskip}
. gen double optiv = Dpystacked_1 - Dpystacked_h_1
{\smallskip}
. ivreg Y1_resid (dtilde=optiv), robust
{\smallskip}
Instrumental variables 2SLS regression          Number of obs     =      2,217
                                                F(1, 2215)        =     110.49
                                                Prob > F          =     0.0000
                                                R-squared         =     0.0888
                                                Root MSE          =       .965
{\smallskip}
\HLI{13}{\TOPT}\HLI{64}
             {\VBAR}               Robust
    Y1_resid {\VBAR} Coefficient  std. err.      t    P>|t|     [95\% conf. interval]
\HLI{13}{\PLUS}\HLI{64}
      dtilde {\VBAR}  -.1075272   .0102295   -10.51   0.000    -.1275875   -.0874668
       _cons {\VBAR}   .0064207   .0205008     0.31   0.754    -.0337821    .0466235
\HLI{13}{\BOTT}\HLI{64}
Instrumented: dtilde
 Instruments: optiv

\end{stlog} 

%

\section{Conclusion}\label{sec:conclusino}
This article introduces the Stata program \texttt{ddml} for Double/Debiased Machine Learning. The program allows for flexible estimation of structural parameters in five econometric models, leveraging a wide range of supervised machine learners. While \texttt{ddml} is compatible with many existing machine learning programs in Stata, it is specifically designed to be used with \texttt{pystacked}, which allows combining several learners into a meta learner via stacking. We see several avenues for extensions: First, \texttt{ddml} primarily focuses on cross-sectional models. Some panel models are readily implementable in \texttt{ddml}, but expanding its capabilities for seamless use with a wide range of panel models would increase its practical relevance. Second, researchers and policymakers are frequently interested in learning treatment effects for specific sub-populations sharing observable characteristics. The estimation of conditional average treatment effects would be a natural extension to the existing \texttt{ddml} program. Third, \texttt{ddml} currently lacks under-identification diagnostics and weak-identification robust inference for instrumental variable regressions, which we hope to add in future releases.

\section{Acknowledgments}
We thank users who tested earlier versions of the program. We also thank Jan Ditzen, Ben Jann, Eroll Kuhn, Di Liu and Moritz Marbach for helpful comments, as well as participants at Stata User Conferences in Germany (2021), Italy (2022), Switzerland (2022) and the UK (2023). All remaining errors are our own.

\bibliographystyle{sj}
\bibliography{library_sj}

\begin{aboutauthors}
Achim Ahrens is Post-Doctoral Researcher and Senior Data Scientist at the Public Policy Group and Immigration Policy Lab, ETH Z\"urich.

Christian B. Hansen is the Wallace W. Booth Professor of Econometrics and Statistics at the University of Chicago Booth School of Business.

Mark E. Schaffer is Professor of Economics in Edinburgh Business School at Heriot-Watt University, Edinburgh, UK, and a Research Fellow at the Institute for the Study of Labour (IZA), Bonn.

Thomas Wiemann is an economics PhD student at the University of Chicago. 
\end{aboutauthors}



\appendix
\pagestyle{empty}
\renewcommand{\thesection}{\Alph{section}}
\renewcommand{\thefigure}{\Alph{section}.\arabic{figure}}
\setcounter{figure}{0}
\renewcommand{\thetable}{\Alph{section}.\arabic{table}}
\setcounter{table}{0}

\newcommand{\bracketr}{]}
\newcommand{\bracketl}{[}

\section{Algorithms}\label{app:algorithms}

\begin{sttech}[Algorithm A.1. DDML for the Interactive Model.]
\noindent Split the sample $\{(Y_i, D_i, \bm{X}_i)\}_{i\in I}$ with $I=\{1,\ldots,n\}$ in $K$ folds of approximately equal size. Denote $I_k$ the set of observations included in fold $k$ and $I_k^c=I\setminus I_k$ its complement.
\begin{enumerate}[nosep] 
    \item Estimate conditional expectations. For each $k$:
    \begin{enumerate}
            \item Fit the CEF estimator to observations in the sub-sample $I_k^c$ for which $D_i=1$ using $Y_i$ as the outcome and $\bm{X}_i$ as predictors to estimate the conditional expectation $g_0(1,\bm{X})=E[Y\vert \bm{X},D=1]$. Obtain the out-of-sample predicted values $\hat{g}_{I_{k_i}^c}(1,\bm{X}_i)$ for $i\in I_k$. Proceed in the same way to obtain $\hat{g}(0,\bm{X})$. 
            \item For each $k$, fit the CEF estimator to the sub-sample $I_k^c$ using $D_i$ as outcome and $\bm{X}_i$ as predictors to estimate the conditional expectation $m(\bm{X})=E[D\vert \bm{X}]$. Obtain the out-of-sample predicted values $\hat{m}_{I_{k_i}^c}(\bm{X}_i)$ for $i\in I_k$.
    \end{enumerate}
    \item Compute the ATE and ATET using (7) and (8).
\end{enumerate}
\end{sttech}

\begin{sttech}[Algorithm A.2. DDML for the Partially Linear IV Model.]
\noindent Split the sample $\{(Y_i, D_i, \bm{X}_i)\}_{i\in I}$ with $I=\{1,\ldots,n\}$ in $K$ folds of approximately equal size. Denote $I_k$ the set of observations included in fold $k$ and $I_k^c=I\setminus I_k$ its complement.
\begin{enumerate}[nosep] 
    \item Estimate conditional expectations. For each $k$:
    \begin{enumerate}
            \item Fit the CEF estimator to the sub-sample $I_k^c$ using $Y_i$ as outcome and $\bm{X}_i$ as predictors to estimate the conditional expectation $\ell_0(\bm{X})=E[Y\vert \bm{X}]$. Obtain the out-of-sample predicted values $\hat{\ell}_{I_{k_i}^c}(\bm{X}_i)$ for $i\in I_k$.
            \item Fit the CEF estimator to the sub-sample $I_k^c$ using $D_i$ as outcome and $\bm{X}_i$ as predictors to estimate the conditional expectation $m_0(\bm{X})=E[D\vert \bm{X}]$. Obtain the out-of-sample predicted values $\hat{m}_{I_{k_i}^c}(\bm{X}_i)$ for $i\in I_k$.
            \item Fit the CEF estimator to the sub-sample $I_k^c$ using $Z_i$ as outcome and $\bm{X}_i$ as predictors to estimate the conditional expectation $r_0(\bm{X})=E[Z\vert \bm{X}]$. Obtain the out-of-sample predicted values $\hat{r}_{I_{k_i}^c}(\bm{X}_i)$ for $i\in I_k$.
    \end{enumerate}
    \item Compute (10).
\end{enumerate}
\end{sttech}

\begin{sttech}[Algorithm A.3. DDML for the Interactive IV Model.]
\noindent Split the sample $\{(Y_i, D_i, \bm{X}_i)\}_{i\in I}$ with $I=\{1,\ldots,n\}$ in $K$ folds of approximately equal size. Denote $I_k$ the set of observations included in fold $k$ and $I_k^c=I\setminus I_k$ its complement.
\begin{enumerate}[nosep] 
    \item Estimate conditional expectations. For each $k$:
    \begin{enumerate}
            \item Fit the CEF estimator to observations in the sub-sample $I_k^c$ for which $Z_i=1$ using $Y_i$ as outcome and $\bm{X}_i$ as predictors to estimate the conditional expectation $\ell_0(1,\bm{X})=E[Y\vert \bm{X},Z=1]$. Obtain the out-of-sample predicted values $\hat{\ell}_{I_{k_i}^c}(1,\bm{X}_i)$ for $i\in I_k$. Proceed in the same way for the estimation of $\ell_0(0,\bm{X})$. 
            \item Fit the CEF estimator to observations in the sub-sample $I_k^c$ for which $Z_i=1$ using $D_i$ as outcome and $\bm{X}_i$ as predictors to estimate the conditional expectation $p_0(1,\bm{X})=\textrm{Pr}(D=1\vert \bm{X},Z=1)$. Obtain the out-of-sample predicted values $\hat{p}_{I_{k_i}^c}(1,\bm{X}_i)$ for $i\in I_k$. Proceed in the same way for the estimation of $p_0(0,\bm{X})$. 
            \item Fit the CEF estimator to the sub-sample $I_k^c$ using $D_i$ as outcome and $\bm{X}_i$ as predictors to estimate the conditional expectation $r(\bm{X})=E[Z\vert \bm{X}]$. Obtain the out-of-sample predicted values $\hat{r}_{I_{k_i}^c}(\bm{X}_i)$ for $i\in I_k$.
    \end{enumerate}
    \item Compute (17).
\end{enumerate}
\end{sttech}

\begin{sttech}[Algorithm A.4. DDML with short-stacking for the Partially Linear Model.]
\noindent Split the sample $\{(Y_i, D_i, \bm{X}_i)\}_{i\in I}$ with $I=\{1,\ldots,n\}$ in $K$ folds of approximately equal size. Denote $I_k$ the set of observations included in fold $k$ and $I_k^c=I\setminus I_k$ its complement. Select a set of $J$ base learners with $J\geq 2$.
\begin{enumerate}[nosep] 
    \item Estimate conditional expectations. For each $k$ and base learner $j$:
    \begin{enumerate}
            \item Fit a CEF estimator $j$ to the sub-sample $I_k^c$ using $Y_i$ as the outcome and $\bm{X}_i$ as predictors. Obtain the out-of-sample predicted values $\hat{\ell}^{(j)}_{I^c_k}(\bm{X}_i)$ for $i\in I_k$.
            \item Fit a CEF estimator $j$  to the sub-sample $I_k^c$ using $D_i$ as the outcome and $\bm{X}_i$ as predictors. Obtain the out-of-sample predicted values $\hat{m}^{(j)}_{I^c_k}(\bm{X}_i)$ for $i\in I_k$.
    \end{enumerate}
    \item Short-stacking:
    \begin{enumerate}
    \item Apply constrained regression of $Y_i$ against $\hat{\ell}^{(1)}_{I^c_k}(\bm{X}_i),\ldots,\hat{\ell}^{(J)}_{I^c_k}(\bm{X}_i)$ using the full sample $I$, which yields the short-stacked predicted values $\hat{\ell}^*(\bm{X}_i)$. 
    \item Apply constrained regression of $D_i$ against $\hat{m}^{(1)}_{I^c_k}(\bm{X}_i),\ldots,\hat{m}^{(J)}_{I^c_k}(\bm{X}_i)$ using the full sample $I$, which yields the short-stacked predicted values $\hat{m}^*(\bm{X}_i)$. 
    \end{enumerate}
    \item Compute the short-stacked DDML esitmator using 
    \begin{align*} 
        \hat{\theta}_n = \frac{\frac{1}{n}\sum_{i=1}^n \big(Y_i-\hat{\ell}^*(\bm{X}_i)\big)\big(D_i-\hat{m}^*(\bm{X}_i)\big)}{\frac{1}{n}\sum_{i=i}^n \big(D_i-\hat{m}^*(\bm{X}_i)\big)^2}.
\end{align*}
\end{enumerate}
\end{sttech}

\begin{sttech}[Algorithm A.5. DDML with short-stacking for the Flexible IV Model.]
\noindent Split the sample $\{(Y_i, D_i, \bm{X}_i)\}_{i\in I}$ with $I=\{1,\ldots,n\}$ in $K$ folds of approximately equal size. Denote $I_k$ the set of observations included in fold $k$ and $I_k^c=I\setminus I_k$ its complement. Select a set of $J$ base learners with $J\geq 2$.
\begin{enumerate}[nosep] 
    \item Estimating conditional expectations $\ell_0(\bm{X})=E[Y\vert \bm{X}]$:
    \begin{enumerate}
            \item For each $k$ and base learner $j$, fit the CEF estimator to the sub-sample $I_k^c$ using $Y_i$ as outcome and $\bm{X}_i$ as predictors. Obtain the out-of-sample predicted values $\hat\ell^{(j)}_{I^c_k}(\bm{X}_i)$ for $i\in I_k$.
            \item Fit a constrained regression of $Y_i$ against $\hat{\ell}^{(j)}_{I^c_k}(\bm{X}_i),\ldots,\hat{\ell}^{(J)}_{I^c_k}(\bm{X}_i)$ over the full sample $I$. The fitted values are the short-stacking estimates $\hat{\ell}^\star(\bm{X}_i)$.
    \end{enumerate}
    \item Estimating conditional expectations $p_0(\bm{X},\bm{Z})=E[D\vert \bm{X},\bm{Z}]$:
    \begin{enumerate}
            \item For each $k$ and base learner $j$, fit the CEF estimator to the sub-sample $I_k^c$ using $D_i$ as outcome and $(\bm{X}_i,\bm{Z}_i)$ as predictors. Obtain the out-of-sample predicted values $\hat{p}^{(j)}_{I^c_k}(\bm{X}_i,\bm{Z}_i)$ for $i\in I_k$, and the in-sample predicted values $\tilde{p}^{(j)}_k(\bm{X}_i,\bm{Z}_i)\equiv\hat{p}^{(j)}_{I^c_k}(\bm{X}_i,\bm{Z}_i)$ for $i\in I_k^c$.
            \item For each $k$, fit a constrained regression of $D_i$ against in-sample predicted values $\tilde{p}^{(1)}_k(\bm{X}_i,\bm{Z}_i),\ldots,\tilde{p}^{(J)}_k(\bm{X}_i,\bm{Z}_i)$ over the sample $I_k^c$ to obtain the out-of-sample short-stack predicted values $\hat{p}^{\star}_{I^c_k}(\bm{X}_i,\bm{Z}_i)$ for $i\in I_k$.
            \item Fit a constrained regression of $D_i$ against $\hat{p}^{(1)}_{I^c_k} (\bm{X}_i,\bm{Z}_i),\ldots,\hat{p}^{(J)}_{I^c_k}(\bm{X}_i,\bm{Z}_i)$ over the full sample $I$. The fitted values are the short-stacking estimates $\hat{p}^\star(\bm{Z}_i, \bm{X}_i)$.
    \end{enumerate}
    \item Estimating conditional expectations $m_0(\bm{X})=E[D\vert \bm{X}]$:
    \begin{enumerate}
            \item For each $k$ and base learner $j$, fit the CEF estimator to the sub-sample $I_k^c$ using in-sample fitted values $\tilde{p}^{(j)}_k(\bm{X}_i,\bm{Z}_i)$ as the outcome and $\bm{X}_i$ as predictors. Obtain the out-of-sample predicted values $\hat{m}^{(j)}_{I^c_k}(\bm{X}_i)$ for $i\in I_k$.
            \item Apply a constrained regression of $\hat{p}_{I^c_k}^\star(\bm{X}_i,\bm{Z}_i)$ against $\hat{m}^{(1)}_{I^c_k}(\bm{X}_i),\ldots,\hat{m}^{(J)}_{I^c_k}(\bm{X}_i)$  over the full sample $I$. The fitted values are the short-stacking estimates $\hat{m}^\star(\bm{X}_i)$.
    \end{enumerate}
        \item Compute \begin{align} 
        \hat{\theta}_{n} = \frac{\frac{1}{n}\sum_{i=1}^n \big(Y_i-\hat{\ell}^\star(\bm{X}_i)\big)\big(\hat{p}^\star(\bm{Z}_i, \bm{X}_i)-\hat{m}^\star(\bm{X}_i)\big)}{\frac{1}{n}\sum_{i=i}^n \big(D_i-\hat{m}^\star(\bm{X}_i)\big)\big(\hat{p}^\star(\bm{Z}_i, \bm{X}_i)-\hat{m}^\star(\bm{X}_i)\big)}.
\end{align}
\end{enumerate}
\end{sttech}

\newpage\section{Additional simulation results}\label{appendix:moremc}

We briefly summarize the results for DGPs~4 and 5 shown in Table~\ref{tab:mc2}. The stacking weights and MSPEs of the individual learners are reported in Tables~\ref{tab:stacking_weights_appendix} and \ref{tab:mspe_appendix}, respectively. 

The bias of DDML with stacking is relatively robust to the inclusion of additional noisy covariates. For $n=100$, DDML with stacking performs at least as well as feasible full-sample estimators. For $n=1000$, DDML with stacking outperforms OLS and PDS-Lasso, and exhibits a bias that is only slightly above the infeasible oracle estimator.  

\begin{table}[H]
    \revisedcolor
    \centering\scriptsize\singlespacing
\begin{threeparttable}
    \caption{Bias and Coverage Rates in the Linear and Nonlinear DGPs}
    \label{tab:mc2}
        \begin{tabular}{rlccccccccc}
    \toprule
    \midrule
          &       & \multicolumn{4}{c}{DGP 4} & & \multicolumn{4}{c}{DGP 5}  \\
    \cline{3-6} \cline{8-11} \\[-6pt]
     & &  \multicolumn{2}{c}{$n=100$}    &\multicolumn{2}{c}{$n=1000$}      &&  \multicolumn{2}{c}{$n=100$}    &\multicolumn{2}{c}{$n=1000$}    \\
    && MAB & Cov.   &  MAB & Cov.   &&   MAB & Cov.   &  MAB & Cov.      \\ \midrule
          \multicolumn{2}{l}{Full sample:}\\
          \partialinput{1}{5}{Simul/sim_SJ/bias_2.tex}
           \multicolumn{2}{l}{DDML methods:}\\
           \multicolumn{2}{l}{\it ~~Base learners}\\
          \partialinput{6}{19}{Simul/sim_SJ/bias_2.tex}
          \multicolumn{2}{l}{\it ~~Meta learners}\\
          \partialinput{20}{23}{Simul/sim_SJ/bias_2.tex}
    \midrule
    \bottomrule
    \end{tabular}%
  \begin{tablenotes}[para,flushleft]
  \footnotesize
  \item \textit{Notes:} The table reports median absolute bias (MAB) and coverage rate of a  95\% confidence interval (CR). We employ standard errors robust to heteroskedasticity.
  For comparison, we report the following full sample estimators: infeasible Oracle, OLS, PDS-Lasso with base and two different expanded sets of covariates. 
  DDML estimators use 20 folds for cross-fitting if $n=100$, and 5 folds if $n=1000$. Meta-learning approaches rely on all listed base learners.
  Results are based on 1,000 replications. Results for DGPs 1-3 can be found in Table~\ref{tab:mc1} in the main text.   
  \vspace{2cm}
  \end{tablenotes}
\end{threeparttable}
\end{table}

\begin{table}[H]
    \revisedcolor
    \vspace*{.9\textheight}\scriptsize
    \begin{rotate}{90}
        \begin{threeparttable}
            \caption{Stacking weights for the estimation of conditional expectation functions}
            \label{tab:stacking_weights_appendix}
            \begin{tabular}{llcccccccccc}
            \toprule
            \midrule
                  &       & \multicolumn{2}{c}{DGP 1} &   \multicolumn{2}{c}{DGP 2} &    \multicolumn{2}{c}{DGP 3}  &    \multicolumn{2}{c}{DGP 4}  &    \multicolumn{2}{c}{DGP 5} \\
             &Observations $n$ & 100 & 1000 &100 & 1000 & 100 & 1000 & 100 & 1000 &100 & 1000 \\  \midrule
                  \multicolumn{2}{l}{\it Estimation of $E[Y\vert \bm{X}]$:}\\
                  &OLS&0.027&0.113&0.016&0.\phantom{000}&0.034&0.002&0.041&0.017&0.156&0.005 \tabularnewline
&Lasso with CV (Base)&0.182&0.642&0.017&0.\phantom{000}&0.206&0.\phantom{000}&0.414&0.016&0.120&0.012 \tabularnewline
&Ridge with CV (Base)&0.350&0.041&0.003&0.\phantom{000}&0.021&0.\phantom{000}&0.033&0.014&0.053&0.018 \tabularnewline
&Lasso with CV (Poly 5)&0.058&0.057&0.205&0.045&0.058&0.\phantom{000}&0.149&0.544&0.099&0.272 \tabularnewline
&Ridge with CV (Poly 5)&0.050&0.042&0.121&0.009&0.014&0.\phantom{000}&0.023&0.001&0.079&0.050 \tabularnewline
&Lasso with CV (Poly 2 + Inter.)&0.037&0.024&0.222&0.939&0.070&0.008&0.074&0.311&0.092&0.343 \tabularnewline
&Ridge with CV (Poly 2 + Inter.)&0.097&0.011&0.071&0.001&0.039&0.\phantom{000}&0.054&0.001&0.043&0.090 \tabularnewline
&Random forest (Low regularization)&0.031&0.004&0.097&0.\phantom{000}&0.083&0.064&0.029&0.\phantom{000}&0.071&0.054 \tabularnewline
&Random forest (Medium regularization)&0.018&0.\phantom{000}&0.066&0.\phantom{000}&0.079&0.\phantom{000}&0.022&0.\phantom{000}&0.041&0.003 \tabularnewline
&Random forest (High regularization)&0.002&0.\phantom{000}&0.021&0.\phantom{000}&0.018&0.\phantom{000}&0.002&0.\phantom{000}&0.012&0.\phantom{000} \tabularnewline
&Gradient boosting (Low regularization)&0.055&0.023&0.058&0.002&0.183&0.156&0.077&0.046&0.054&0.033 \tabularnewline
&Gradient boosting (Medium regularization)&0.021&0.013&0.038&0.\phantom{000}&0.124&0.687&0.033&0.023&0.021&0.013 \tabularnewline
&Gradient boosting (High regularization)&0.003&0.\phantom{000}&0.022&0.\phantom{000}&0.025&0.071&0.003&0.\phantom{000}&0.006&0.\phantom{000} \tabularnewline
&Neural net&0.069&0.029&0.045&0.004&0.047&0.012&0.045&0.026&0.154&0.105 \tabularnewline
 
                  \\
                   \multicolumn{2}{l}{\it Estimation of $E[D\vert \bm{X}]$:}\\
                  &OLS&0.029&0.119&0.017&0.\phantom{000}&0.034&0.002&0.044&0.016&0.161&0.006 \tabularnewline
&Lasso with CV (Base)&0.171&0.637&0.016&0.\phantom{000}&0.182&0.\phantom{000}&0.405&0.014&0.123&0.012 \tabularnewline
&Ridge with CV (Base)&0.350&0.034&0.003&0.\phantom{000}&0.020&0.\phantom{000}&0.031&0.013&0.053&0.018 \tabularnewline
&Lasso with CV (Poly 5)&0.059&0.058&0.218&0.045&0.057&0.\phantom{000}&0.151&0.561&0.103&0.272 \tabularnewline
&Ridge with CV (Poly 5)&0.048&0.047&0.116&0.006&0.014&0.\phantom{000}&0.026&0.001&0.074&0.049 \tabularnewline
&Lasso with CV (Poly 2 + Inter.)&0.038&0.024&0.237&0.943&0.068&0.007&0.076&0.301&0.099&0.340 \tabularnewline
&Ridge with CV (Poly 2 + Inter.)&0.094&0.011&0.067&0.001&0.039&0.\phantom{000}&0.054&0.001&0.044&0.096 \tabularnewline
&Random forest (Low regularization)&0.036&0.005&0.085&0.\phantom{000}&0.083&0.047&0.030&0.\phantom{000}&0.066&0.055 \tabularnewline
&Random forest (Medium regularization)&0.022&0.\phantom{000}&0.063&0.\phantom{000}&0.079&0.\phantom{000}&0.022&0.\phantom{000}&0.036&0.003 \tabularnewline
&Random forest (High regularization)&0.002&0.\phantom{000}&0.021&0.\phantom{000}&0.024&0.\phantom{000}&0.003&0.\phantom{000}&0.015&0.\phantom{000} \tabularnewline
&Gradient boosting (Low regularization)&0.055&0.024&0.055&0.001&0.190&0.171&0.079&0.050&0.054&0.034 \tabularnewline
&Gradient boosting (Medium regularization)&0.022&0.012&0.036&0.\phantom{000}&0.139&0.731&0.032&0.020&0.022&0.012 \tabularnewline
&Gradient boosting (High regularization)&0.004&0.\phantom{000}&0.021&0.\phantom{000}&0.024&0.030&0.003&0.\phantom{000}&0.006&0.\phantom{000} \tabularnewline
&Neural net&0.069&0.029&0.046&0.003&0.045&0.011&0.045&0.023&0.146&0.103 \tabularnewline
 \\[-.3cm]
            \midrule
            \bottomrule
            \end{tabular}%
          \begin{tablenotes}[para,flushleft]
          \footnotesize
          \item \textit{Notes:} This table shows the stacking weights averaged over bootstrap and cross-fitting iterations for each base learner. We report the stacking weights for the estimation of $E[Y\vert \bm{X}]$ and $E[D\vert \bm{X}]$, and for sample sizes of $n=100$ and $n=1000$.
          \end{tablenotes}
            \end{threeparttable}
    \end{rotate}
\end{table}
 
\begin{table}[H]
    \vspace*{.9\textheight}\scriptsize
    \begin{rotate}{90}
        \begin{threeparttable}
            \caption{Mean-squared prediction error the estimation of conditional expectation functions}
            \label{tab:mspe_appendix}
            \begin{tabular}{llcccccccccc}
            \toprule
            \midrule
                  &       & \multicolumn{2}{c}{DGP 1} &   \multicolumn{2}{c}{DGP 2} &    \multicolumn{2}{c}{DGP 3}  &    \multicolumn{2}{c}{DGP 4}  &    \multicolumn{2}{c}{DGP 5} \\
             &Observations $n$ & 100 & 1000 &100 & 1000 & 100 & 1000 & 100 & 1000 &100 & 1000 \\  \midrule
                  \multicolumn{2}{l}{\it Estimation of $E[Y\vert \bm{X}]$:}\\
                  &OLS&2.753&1.372&4.625&2.085&3.292&1.165&3.364&1.766&1.811&1.687 \tabularnewline
&Lasso with CV (Base)&0.941&1.148&1.900&1.882&0.801&0.909&1.299&1.492&1.572&1.611 \tabularnewline
&Ridge with CV (Base)&0.760&1.258&1.887&1.875&0.761&0.752&1.147&1.765&1.661&1.687 \tabularnewline
&Lasso with CV (Poly 5)&1.254&1.030&2.783&2.608&0.988&0.951&1.575&1.575&1.926&1.742 \tabularnewline
&Ridge with CV (Poly 5)&1.176&0.940&2.445&2.471&0.777&0.849&1.468&1.304&1.591&1.764 \tabularnewline
&Lasso with CV (Poly 2 + Inter.)&0.545&0.872&2.119&2.614&0.728&0.916&1.096&1.450&1.534&1.677 \tabularnewline
&Ridge with CV (Poly 2 + Inter.)&3.299&0.787&2.888&2.520&1.392&0.562&2.629&1.247&1.203&1.904 \tabularnewline
&Random forest (Low regularization)&0.288&0.492&2.016&2.026&0.771&0.985&0.927&1.113&1.327&1.512 \tabularnewline
&Random forest (Medium regularization)&0.272&0.378&1.991&1.922&0.752&0.816&0.907&0.975&1.290&1.350 \tabularnewline
&Random forest (High regularization)&0.128&0.110&1.866&1.840&0.567&0.529&0.725&0.698&0.925&0.852 \tabularnewline
&Gradient boosting (Low regularization)&0.959&1.040&2.458&2.478&1.568&1.440&1.600&1.623&1.806&1.729 \tabularnewline
&Gradient boosting (Medium regularization)&0.581&0.778&2.118&2.226&1.121&1.260&1.220&1.407&1.392&1.498 \tabularnewline
&Gradient boosting (High regularization)&0.183&0.317&1.898&1.890&0.699&0.962&0.818&0.995&0.928&1.037 \tabularnewline
&Neural net&1.614&2.116&2.925&3.469&1.856&2.326&2.035&2.612&1.983&1.944 \tabularnewline
 
                  \\
                   \multicolumn{2}{l}{\it Estimation of $E[D\vert \bm{X}]$:}\\
                  &OLS&2.166&1.063&3.801&1.737&2.615&0.925&2.673&1.422&1.458&1.364 \tabularnewline
&Lasso with CV (Base)&0.733&0.886&1.584&1.571&0.641&0.721&1.043&1.201&1.266&1.303 \tabularnewline
&Ridge with CV (Base)&0.590&0.950&1.573&1.565&0.609&0.595&0.915&1.421&1.342&1.364 \tabularnewline
&Lasso with CV (Poly 5)&1.008&0.793&2.274&2.182&0.966&0.751&1.404&1.271&1.497&1.407 \tabularnewline
&Ridge with CV (Poly 5)&0.953&0.729&2.077&2.061&0.629&0.674&1.220&1.050&1.248&1.423 \tabularnewline
&Lasso with CV (Poly 2 + Inter.)&0.420&0.671&1.776&2.192&0.586&0.726&0.880&1.170&1.232&1.357 \tabularnewline
&Ridge with CV (Poly 2 + Inter.)&2.541&0.615&2.415&2.114&1.110&0.443&2.100&1.002&0.967&1.540 \tabularnewline
&Random forest (Low regularization)&0.221&0.376&1.659&1.680&0.596&0.763&0.725&0.881&1.056&1.209 \tabularnewline
&Random forest (Medium regularization)&0.210&0.295&1.642&1.600&0.583&0.642&0.712&0.780&1.027&1.087 \tabularnewline
&Random forest (High regularization)&0.100&0.087&1.549&1.536&0.445&0.420&0.576&0.563&0.742&0.690 \tabularnewline
&Gradient boosting (Low regularization)&0.732&0.791&2.005&2.041&1.220&1.137&1.250&1.291&1.424&1.378 \tabularnewline
&Gradient boosting (Medium regularization)&0.440&0.583&1.739&1.841&0.884&0.997&0.951&1.118&1.101&1.191 \tabularnewline
&Gradient boosting (High regularization)&0.129&0.205&1.573&1.569&0.554&0.733&0.635&0.767&0.723&0.796 \tabularnewline
&Neural net&1.290&1.673&2.428&2.886&1.491&1.867&1.637&2.111&1.608&1.578 \tabularnewline
 \\[-.3cm]
            \midrule
            \bottomrule
            \end{tabular}%
          \begin{tablenotes}[para,flushleft]
          \footnotesize
          \item \textit{Notes:} This table shows the cross-fitted mean-squared prediction error averaged over bootstrap iterations for the listed conditional expectation function estimators. We report the mean-squared predictor error for the estimation of $E[Y\vert \bm{X}]$ and $E[D\vert \bm{X}]$, and for sample sizes of $n=100$ and $n=1000$.
          \end{tablenotes}
            \end{threeparttable}
    \end{rotate}
\end{table}

\newpage\section{Applications}\label{appendix:moreapplications}

Here we continue the 401(k) application from the main text to illustrate estimation of the interactive model and IV models. We use the same data and variables as outlined in the main text. For the IV models, we use eligibility to enroll for the 401(k) pension plan as the instrument and treat participation in a 401(k) as the endogenous variable.

\subsection{Interactive Model (\texttt{interactive})}\label{sec:interactive_application}

We allow for heterogenous treatment effects using the \texttt{interactive} model. To this end, the conditional expectation of \verb+$Y+ given \verb+$X+ is fit separately for $D=1$ and $D=0$. We also use \texttt{reps(5)}. This will execute the \texttt{ddml} estimation three times using different random folds. This reduces dependence on a specific fold.

\begin{stlog}
. *** initialize
. set seed 123
{\smallskip}
. ddml init interactive, kfolds(5) reps(5)

\end{stlog}

\begin{stlog}
. *** add learners for E[Y|X,D=0] and E[Y|X,D=0]
. ddml E[Y|X,D]: pystacked $Y $X                                           || ///
>     method(ols)                                                          || ///
>     m(lassocv) xvars(c.($X)\#\#c.($X))                                     || ///
>     m(ridgecv) xvars(c.($X)\#\#c.($X))                                     || ///
>     m(rf) pipe(sparse) opt(max_features(5))                              || ///
>     m(gradboost) pipe(sparse) opt(n_estimators(250) learning_rate(0.01)) ,  ///
>     njobs(5)
Learner Y1_pystacked added successfully.
{\smallskip}
. *** add learners for E[D|X]
. ddml E[D|X]: pystacked $D $X                                             || ///
>     method(ols)                                                          || ///
>     m(lassocv) xvars(c.($X)\#\#c.($X))                                     || ///
>     m(ridgecv) xvars(c.($X)\#\#c.($X))                                     || ///
>     m(rf) pipe(sparse) opt(max_features(5))                              || ///
>     m(gradboost) pipe(sparse) opt(n_estimators(250) learning_rate(0.01)) ,  ///
>     njobs(5)
Learner D1_pystacked added successfully.

\end{stlog}

\begin{stlog}
. ddml estimate
{\smallskip}
{\smallskip}
Model:                  interactive, crossfit folds k=5, resamples r=5
Mata global (mname):    m0
Dependent variable (Y): net_tfa
 net_tfa learners:      Y1_pystacked
D equations (1):        e401
 e401 learners:         D1_pystacked
{\smallskip}
DDML estimation results (ATE):
spec  r  Y(0) learner  Y(1) learner     D learner         b        SE 
  st  1  Y1_pystacked  Y1_pystacked  D1_pystacked  8026.894 (1126.459)
  st  2  Y1_pystacked  Y1_pystacked  D1_pystacked  7879.717 (1122.815)
  st  3  Y1_pystacked  Y1_pystacked  D1_pystacked  8050.997 (1119.537)
  st  4  Y1_pystacked  Y1_pystacked  D1_pystacked  8157.737 (1113.299)
  st  5  Y1_pystacked  Y1_pystacked  D1_pystacked  7753.948 (1138.377)
{\smallskip}
Mean/med Y(0) learner  Y(1) learner     D learner         b        SE 
  st mn  Y1_pystacked  Y1_pystacked  D1_pystacked  7973.859 (1132.658)
  st md  Y1_pystacked  Y1_pystacked  D1_pystacked  8026.894 (1126.459)
{\smallskip}
Median over 5 stacking resamples (ATE)
E[y|X,D=0]   = Y1_pystacked0                       Number of obs   =      9915
E[y|X,D=1]   = Y1_pystacked1
E[D|X]       = D1_pystacked
\HLI{13}{\TOPT}\HLI{64}
             {\VBAR}               Robust
     net_tfa {\VBAR} Coefficient  std. err.      z    P>|z|     [95\% conf. interval]
\HLI{13}{\PLUS}\HLI{64}
        e401 {\VBAR}   8026.894   1126.459     7.13   0.000     5819.075    10234.71
\HLI{13}{\BOTT}\HLI{64}
Stacking final estimator: nnls1
Warning: 5 resamples had propensity scores trimmed to lower limit .01.
{\smallskip}
Summary over 5 resamples:
       D eqn      mean       min       p25       p50       p75       max
        e401   7973.8585 7753.9478 7879.7173 8026.8940 8050.9966 8157.7368
{\smallskip}

\end{stlog}

\paragraph{One-line syntax (output omitted).}

\begin{stlog}
. qui qddml $Y $D ($X), model(interactive)
{\smallskip}

\end{stlog}

\subsection{IV model (\texttt{iv})}

\begin{stlog}
. use "sipp1991.dta", clear
{\smallskip}
. global Y net_tfa
{\smallskip}
. global X age inc educ fsize marr twoearn db pira hown
{\smallskip}
. global Z e401
{\smallskip}
. global D p401
{\smallskip}

\end{stlog}

\subsubsection*{Step 1: Initialize \texttt{ddml} model.}
\begin{stlog}
. set seed 123
{\smallskip}
. ddml init iv

\end{stlog}

\subsubsection*{Step 2: Add supervised machine learners for estimating conditional expectations.}
\begin{stlog}
. *** E[Y|X]
. ddml E[Y|X]: pystacked $Y $X                                             || ///
>     method(ols)                                                          || ///
>     m(lassocv) xvars(c.($X)\#\#c.($X))                                     || ///
>     m(ridgecv) xvars(c.($X)\#\#c.($X))                                     || ///
>     m(rf) pipe(sparse) opt(max_features(5))                              || ///
>     m(gradboost) pipe(sparse) opt(n_estimators(250) learning_rate(0.01)) ,  ///
>     njobs(4)
Learner Y1_pystacked added successfully.
{\smallskip}
. *** E[D|X]
. ddml E[D|X]: pystacked $D $X                                             || ///
>     method(ols)                                                          || ///
>     m(lassocv) xvars(c.($X)\#\#c.($X))                                     || ///
>     m(ridgecv) xvars(c.($X)\#\#c.($X))                                     || ///
>     m(rf) pipe(sparse) opt(max_features(5))                              || ///
>     m(gradboost) pipe(sparse) opt(n_estimators(250) learning_rate(0.01)) ,  ///
>     njobs(4)
Learner D1_pystacked added successfully.
{\smallskip}
. *** E[Z|X]
. ddml E[Z|X]: pystacked $Z $X                                             || ///
>     method(ols)                                                          || ///
>     m(lassocv) xvars(c.($X)\#\#c.($X))                                     || ///
>     m(ridgecv) xvars(c.($X)\#\#c.($X))                                     || ///
>     m(rf) pipe(sparse) opt(max_features(5))                              || ///
>     m(gradboost) pipe(sparse) opt(n_estimators(250) learning_rate(0.01)) ,  ///
>     njobs(4)
Learner Z1_pystacked added successfully.

\end{stlog}

\subsubsection*{Step 3: Perform cross-fitting.}
\begin{stlog}
. ddml crossfit
Cross-fitting E[y|X] equation: net_tfa
Cross-fitting fold 1 2 3 4 5 ...completed cross-fitting
Cross-fitting E[D|X] equation: p401
Cross-fitting fold 1 2 3 4 5 ...completed cross-fitting
Cross-fitting E[Z|X]: e401
Cross-fitting fold 1 2 3 4 5 ...completed cross-fitting
{\smallskip}
. ddml extract, show(pystacked)
{\smallskip}
pystacked weights for Z1_pystacked (e401)
             learner   resample     fold_1     fold_2     fold_3     fold_4     fold_5
      ols          1          1          0          0  1.577e-17          0  .04870791
  lassocv          2          1          0  .36068464  .19641246  .23065829          0
  ridgecv          3          1  .50644273  .32367528  .36796975  .30346779  .45676698
       rf          4          1  .04990979  .03383132  8.761e-18  .04246409  .02379444
gradboost          5          1  .44364749  .28180876  .43561779  .42340983  .47073066
{\smallskip}
mean stacking weights across folds/resamples for Z1_pystacked (e401)
final stacking estimator: nnls1
               learner  mean_weight        rep_1
      ols            1    .00974158    .00974158
  lassocv            2    .15755108    .15755108
  ridgecv            3     .3916645     .3916645
       rf            4    .02999993    .02999993
gradboost            5     .4110429     .4110429
{\smallskip}
pystacked MSEs for Z1_pystacked (e401)
             learner   resample     fold_1     fold_2     fold_3     fold_4     fold_5
      ols          1          1  .19946988  .20137585  .20302906  .20030431   .2013705
  lassocv          2          1  .19523544  .19689361  .19807185  .19569792  .19799156
  ridgecv          3          1  .19512701  .19728582  .19849531  .19614749   .1981114
       rf          4          1  .21307726    .215616  .21841171  .21446185  .21691469
gradboost          5          1  .19529958  .19797459  .19862763  .19615183  .19795431
{\smallskip}
mean stacking MSEs across folds/resamples for Z1_pystacked (e401)
final stacking estimator: nnls1
             learner   mean_MSE      rep_1
      ols          1  .20110992  .20110992
  lassocv          2  .19677808  .19677808
  ridgecv          3  .19703341  .19703341
       rf          4   .2156963   .2156963
gradboost          5  .19720159  .19720159
{\smallskip}
pystacked weights for Y1_pystacked (net_tfa)
             learner   resample     fold_1     fold_2     fold_3     fold_4     fold_5
      ols          1          1  .15175107  .01201745  .14501553  .07072362  .03857199
  lassocv          2          1          0  .86668516          0          0  .26575544
  ridgecv          3          1  .76036209  .10111911  .80863601  .75672014  .61229963
       rf          4          1   .0802917          0   .0208563          0  .06587322
gradboost          5          1          0  .01161239  .02847589  .19812117  .01749972
{\smallskip}
mean stacking weights across folds/resamples for Y1_pystacked (net_tfa)
final stacking estimator: nnls1
               learner  mean_weight        rep_1
      ols            1    .08361593    .08361593
  lassocv            2    .22648812    .22648812
  ridgecv            3    .60782739    .60782739
       rf            4    .03340424    .03340424
gradboost            5    .05114183    .05114183
{\smallskip}
pystacked MSEs for Y1_pystacked (net_tfa)
             learner   resample     fold_1     fold_2     fold_3     fold_4     fold_5
      ols          1          1  3.135e+09  3.040e+09  2.674e+09  3.438e+09  3.378e+09
  lassocv          2          1  2.958e+09  2.843e+09  2.463e+09  3.107e+09  3.025e+09
  ridgecv          3          1  2.952e+09  2.857e+09  2.434e+09  3.099e+09  3.024e+09
       rf          4          1  3.344e+09  3.213e+09  2.765e+09  3.541e+09  3.493e+09
gradboost          5          1  3.217e+09  2.990e+09  2.656e+09  3.317e+09  3.344e+09
{\smallskip}
mean stacking MSEs across folds/resamples for Y1_pystacked (net_tfa)
final stacking estimator: nnls1
             learner   mean_MSE      rep_1
      ols          1  3.133e+09  3.133e+09
  lassocv          2  2.879e+09  2.879e+09
  ridgecv          3  2.873e+09  2.873e+09
       rf          4  3.271e+09  3.271e+09
gradboost          5  3.105e+09  3.105e+09
{\smallskip}
pystacked weights for D1_pystacked (p401)
             learner   resample     fold_1     fold_2     fold_3     fold_4     fold_5
      ols          1          1  .05947061  8.397e-18  .06447698  .06941268  .06619485
  lassocv          2          1  .05876314  .32796901  .20842013  .39455972  .21484529
  ridgecv          3          1  .41441662   .2760156  .36272642  .22474909  .47123129
       rf          4          1  .03741096  .05709511  .02323724   .0711886  .08544984
gradboost          5          1  .42993867  .33892028  .34113922  .24008991  .16227873
{\smallskip}
mean stacking weights across folds/resamples for D1_pystacked (p401)
final stacking estimator: nnls1
               learner  mean_weight        rep_1
      ols            1    .05191103    .05191103
  lassocv            2    .24091146    .24091146
  ridgecv            3    .34982781    .34982781
       rf            4    .05487635    .05487635
gradboost            5    .30247336    .30247336
{\smallskip}
pystacked MSEs for D1_pystacked (p401)
             learner   resample     fold_1     fold_2     fold_3     fold_4     fold_5
      ols          1          1  .17102922  .17168453  .17532105  .17229367  .17290828
  lassocv          2          1  .16965393  .17005015  .17383906  .17092387  .17131014
  ridgecv          3          1  .16988454  .17045619  .17415227  .17155958  .17134841
       rf          4          1   .1850616   .1848705  .19065443  .18625053   .1863959
gradboost          5          1  .16981445  .17040097  .17420905  .17155253  .17207733
{\smallskip}
mean stacking MSEs across folds/resamples for D1_pystacked (p401)
final stacking estimator: nnls1
             learner   mean_MSE      rep_1
      ols          1  .17264735  .17264735
  lassocv          2  .17115543  .17115543
  ridgecv          3   .1714802   .1714802
       rf          4  .18664659  .18664659
gradboost          5  .17161087  .17161087
{\smallskip}

\end{stlog}

\subsubsection*{Step 4: Estimate causal effects.}
\begin{stlog}
. ddml estimate 
{\smallskip}
{\smallskip}
Model:                  iv, crossfit folds k=5, resamples r=1
Mata global (mname):    m0
Dependent variable (Y): net_tfa
 net_tfa learners:      Y1_pystacked
D equations (1):        p401
 p401 learners:         D1_pystacked
Z equations (1):        e401
 e401 learners:         Z1_pystacked
{\smallskip}
DDML estimation results:
spec  r     Y learner     D learner         b        SE      Z learner
  st  1  Y1_pystacked  D1_pystacked 13528.537 (1726.023)  Z1_pystacked
{\smallskip}
Stacking DDML model
y-E[y|X]  = y-Y1_pystacked_1                       Number of obs   =      9915
D-E[D|X]  = D-D1_pystacked_1 
Z-E[Z|X]  = Z-Z1_pystacked_1 
\HLI{13}{\TOPT}\HLI{64}
     net_tfa {\VBAR} Coefficient  Std. err.      z    P>|z|     [95\% conf. interval]
\HLI{13}{\PLUS}\HLI{64}
        p401 {\VBAR}   13528.54   1726.023     7.84   0.000     10145.59    16911.48
       _cons {\VBAR}  -42.83654   531.1319    -0.08   0.936    -1083.836    998.1628
\HLI{13}{\BOTT}\HLI{64}
Stacking final estimator: nnls1
{\smallskip}
{\smallskip}

\end{stlog}

\subsubsection*{One-line syntax.}
\begin{stlog}
. qui qddml $Y ($X) ($D = $Z), model(iv)
{\smallskip}

\end{stlog}

\subsection{Interactive IV Model \texttt{interactiveiv}}\label{appendix:late_application}

\begin{stlog}
. use "sipp1991.dta", clear
{\smallskip}
. global Y net_tfa
{\smallskip}
. global X age inc educ fsize marr twoearn db pira hown
{\smallskip}
. global Z e401
{\smallskip}
. global D p401
{\smallskip}

\end{stlog} 

\paragraph{Step 1: Initialize \texttt{ddml} model.}
\begin{stlog}
. set seed 123
{\smallskip}
. ddml init interactiveiv

\end{stlog} 

\paragraph{Step 2: Add supervised machine learners for estimating conditional expectations.}
\begin{stlog}
. * add learners for E[Y|X,Z=0] and E[Y|X,Z=0]
. ddml E[Y|X,Z]: pystacked $Y $X                                           || ///
>     method(ols)                                                          || ///
>     m(lassocv) xvars(c.($X)\#\#c.($X))                                     || ///
>     m(ridgecv) xvars(c.($X)\#\#c.($X))                                     || ///
>     m(rf) pipe(sparse) opt(max_features(5))                              || ///
>     m(gradboost) pipe(sparse) opt(n_estimators(250) learning_rate(0.01)) ,  ///
>     njobs(4)
Learner Y1_pystacked added successfully.
{\smallskip}
. * add learners for E[D|X,Z=0] and E[D|X,Z=1]
. ddml E[D|X,Z]: pystacked $D $X                                           || ///
>     method(ols)                                                          || ///
>     m(lassocv) xvars(c.($X)\#\#c.($X))                                     || ///
>     m(ridgecv) xvars(c.($X)\#\#c.($X))                                     || ///
>     m(rf) pipe(sparse) opt(max_features(5))                              || ///
>     m(gradboost) pipe(sparse) opt(n_estimators(250) learning_rate(0.01)) ,  ///
>     njobs(4)
Learner D1_pystacked added successfully.
{\smallskip}
. * add learners for E[Z|X]
. ddml E[Z|X]: pystacked $Z $X                                             || ///
>     method(ols)                                                          || ///
>     m(lassocv) xvars(c.($X)\#\#c.($X))                                     || ///
>     m(ridgecv) xvars(c.($X)\#\#c.($X))                                     || ///
>     m(rf) pipe(sparse) opt(max_features(5))                              || ///
>     m(gradboost) pipe(sparse) opt(n_estimators(250) learning_rate(0.01)) ,  ///
>     njobs(4)
Learner Z1_pystacked added successfully.

\end{stlog} 

\paragraph{Step 3: Perform cross-fitting.}
\begin{stlog}
. ddml crossfit 
Cross-fitting E[y|X,Z] equation: net_tfa
Cross-fitting fold 1 2 3 4 5 ...completed cross-fitting
Cross-fitting E[D|X,Z] equation: p401
Cross-fitting fold 1 2 3 4 5 ...completed cross-fitting
Cross-fitting E[Z|X]: e401
Cross-fitting fold 1 2 3 4 5 ...completed cross-fitting

\end{stlog} 

\paragraph{Step 4: Estimate causal effects.}
\begin{stlog}
. ddml estimate  
{\smallskip}
{\smallskip}
Model:                  interactiveiv, crossfit folds k=5, resamples r=1
Mata global (mname):    m0
Dependent variable (Y): net_tfa
 net_tfa learners:      Y1_pystacked
D equations (1):        p401
 p401 learners:         D1_pystacked
Z equations (1):        e401
 e401 learners:         Z1_pystacked
{\smallskip}
DDML estimation results (LATE):
spec  r  Y(0) learner  Y(1) learner  D(0) learner  D(1) learner         b        SE      Z learner
  st  1  Y1_pystacked  Y1_pystacked  D1_pystacked  D1_pystacked 11579.166 (1613.058)  Z1_pystacked
{\smallskip}
Stacking DDML model (LATE)
E[y|X,Z=0]   = Y1_pystacked0_1                     Number of obs   =      9915
E[y|X,Z=1]   = Y1_pystacked1_1
E[D|X,Z=0]   = D1_pystacked0_1
E[D|X,Z=1]   = D1_pystacked1_1
E[Z|X]       = Z1_pystacked_1
\HLI{13}{\TOPT}\HLI{64}
             {\VBAR}               Robust
     net_tfa {\VBAR} Coefficient  std. err.      z    P>|z|     [95\% conf. interval]
\HLI{13}{\PLUS}\HLI{64}
        p401 {\VBAR}   11579.17   1613.058     7.18   0.000     8417.631     14740.7
\HLI{13}{\BOTT}\HLI{64}
Stacking final estimator: nnls1
Warning: 13 propensity scores trimmed to lower limit .01.
{\smallskip}

\end{stlog} 

\paragraph{One-line syntax.}
\begin{stlog}
. qui qddml $Y ($X) ($D=$Z), model(interactiveiv)
{\smallskip}

\end{stlog}

\end{document}